%% file: art_20140324arxiv.tex
\documentclass[11pt]{article}

\usepackage{amssymb}
\usepackage{amsmath}
\usepackage{a4wide}

\usepackage{graphicx}
\usepackage[utf8]{inputenc}
 \usepackage{lineno}
 \usepackage{natbib}
\usepackage{ulem}
 \usepackage{xcolor} %couleurs a mettre avant paquet hyperref pour les couleurs
\usepackage{geometry}
   \usepackage{url}
\usepackage{subfig} % to add fig A. fig. B, etc.

\newcommand{\E}{\mathbb{E}}
\renewcommand{\P}{\mathbb{P}}

\title{Network impact on persistence in a finite population
dynamic diffusion model: application to an emergent seed exchange network}
\author{
Pierre Barbillon\footnote{These authors contributed equally to this work.}
\footnote{Corresponding author: 16 rue Claude Bernard,
75231 Paris Cedex 05,
pierre.barbillon@agroparistech.fr}\\
AgroParisTech / UMR INRA MIA,\\ F-75005 Paris, France\\
INRA, UMR 518,\\ F-75005 Paris, France
\and
Mathieu Thomas\footnotemark[1]\\
AgroParisTech / UMR INRA MIA,\\ F-75005 Paris, France\\
INRA, UMR 518,\\ F-75005 Paris, France
\and
Isabelle Goldringer\\
INRA, UMR 0320 / UMR 8120\\ Génétique Végétale,\\  F-91190 Gif-sur-Yvette, France
\and
Frédéric Hospital\\
INRA, UMR 1313\\ Génétique Animale et Biologie Intégrative,\\ F-78352 Jouy-en-Josas, France
\and
Stéphane Robin\\
AgroParisTech / UMR INRA MIA,\\ F-75005 Paris, France\\
INRA, UMR 518,\\ F-75005 Paris, France
}

\date{}
\begin{document}

\maketitle


\begin{abstract}

Dynamic extinction colonisation models (also called contact processes) are widely studied in epidemiology and in metapopulation theory.  
Contacts are usually assumed to be possible only through a network of connected patches. 
This network accounts for a spatial landscape or a social organisation of interactions. 
Thanks to social network literature, heterogeneous networks of contacts can be considered.
A major issue is to assess the influence of the network in the dynamic model.
Most work with this common purpose uses deterministic models or an approximation of a stochastic Extinction-Colonisation model (sEC) which are
relevant only for large networks. When working with a limited size network, the induced stochasticity is essential and has to
be taken into account in the conclusions. 
Here, a rigorous framework is proposed for limited size networks and the limitations of the deterministic approximation are exhibited.
This framework allows exact computations when the number of patches is small. Otherwise, 
simulations are used and enhanced by
adapted simulation techniques when necessary. 
A sensitivity analysis was conducted to compare four main topologies of networks in contrasting settings to determine the
role of the network.
A challenging case was studied in this context: seed exchange of crop species in the R\'eseau Semences Paysannes (RSP),
an emergent French farmers' organisation.
A stochastic Extinction-Colonisation model was used to characterize the consequences of substantial changes in terms of RSP's social organisation on the ability of the system to maintain crop varieties.

\vspace{0.5cm} \noindent \textbf{Keywords:}
metapopulation; social network; finite-population model; sensitivity analysis;
seed exchange network.
\end{abstract}

\section{Introduction}

To deal with the persistence of a metapopulation in a dynamic extinction-colonisation model, 
several studies have used deterministic models where the evolution is described by differential equations \citep[see][]{Levins1969,Hanski2000,sole_self-organization_2006}. 
These models are grounded on an asymptotic approximation in the number of patches. 
The same models are used in epidemiology (SIS: Susceptible Infected Susceptible model).
More recently, some studies have dealt with the stochastic effect due to a finite and limited number of patches/actors.
\citet{Chakrabarti2008} have proposed an approximation in the stochastic model which leads to conclusions similar to the ones obtained with deterministic models.
\citet{Gilarranz201211} have shown by simulations the impact of stochasticity due to a limited number of patches and they
have underscored the differences with the results obtained with deterministic models when comparing the ability of different networks to conserve a metapopulation.
However, their results depend only on the ratio of the extinction rate to the colonisation rate which is not relevant in a stochastic model.
Indeed, the same ratio values with different values of the extinction and colonisation rates can lead to very different situations for the dynamic of the metapopulation.\\

In this paper, we study the same stochastic model as \citet{Gilarranz201211}. The patches can be in only two states: occupied or empty. 
The dynamic consists in a succession of extinction events followed by colonisation events. 
We provide a rigorous theoretical basis to this model which explains the different behaviours observed in the simulations. 
Indeed, the stochastic model is a Markov chain and its transition matrix can be constructed as done by \citet{Day1995}. 
From this, we deduce that there is a unique possible equilibrium which is the absorbing state when all patches are empty.
Moreover, the steady state which can be observed where the number of occupied patches seems to have reached an equilibrium
corresponds to a so-called quasi-stationary distribution of the Markov chain \citep{Darroch1965,Meleard2012}.
On this basis, in order to assess the persistence of a metapopulation, we propose criteria which are adapted to the stochastic context.
In particular, since the metapopulation will become extinct in any case, we decide to fix a limited time-horizon and to provide conclusions relying on this time-horizon.
We also show the limitations of the asymptotic approximation.
Furthermore, this approach leads to exact computations provided that the number of patches is not too large. Otherwise, simulations
can be conducted and enhanced by modified simulation techniques when necessary.
\\

The goal of this study is to measure the impact of the interaction network which 
describes the relationships between patches (during colonisation events) on the behaviour of the dynamic model. 
Following the metapopulation model, 
the network used to account for heterogeneous spatial organisation \citep{Gilarranz201211} can also be used to account for a social organisation \citep{read_dynamic_2008}. 
Indeed, this work was designed in the context of social networks of farmers who exchange seeds, an important social process in the diffusion 
and maintenance of crop biodiversity \citep[reviewed in][]{thomas_seed_2011}. 
We assume that seeds spread through farmers' relationships like an epidemiological process as suggested by \citet{pautasso_seed_2013}.
Relying on the study of the R\'eseau Semences Paysannes (RSP) which is a
French network of farmers involved in seed exchange of heirloom crop species \citep{demeulenaere_etude_2008,demeulenaere_semences_2012,thomas_-farm_2012}, we compare different scenarios of social organisation and attest
their effects on the persistence of one crop variety.

In the following, the network is assumed to be non-oriented and is denoted by $G$.
In the deterministic work \citep{Hanski2000} or
in approximation of the stochastic model \citep{Chakrabarti2008}, the leading eigenvalue of the adjacency matrix of $G$ is sufficient to describe the
impact of the network on persistence. 
We show that this is no longer true in a stochastic model. 
We propose to study four main network topologies which represent really distinct organisations. 
These topologies are determined by generative models:
an Erd\H{o}s-Rényi model \citep{erdosrenyi1959}, a community model (where connection inside a community is more likely than between two patches from different communities), a preferential attachment model \citep{Albert2002} 
and a ``lattice'' model where nodes all have approximately the same degree. 
\\

In section \ref{secModel}, a full description and an analysis of the sEC model are provided together with the algorithms used in simulations.
The limits of the deterministic approximation are presented.
The topologies for networks are detailed in section \ref{secNetwork}.
We conducted a sensitivity analysis to measure the impact of the topology in contrasting settings.
The results are  presented in section \ref{secAnalysisSensitivity}.
A motivating application of this work in section \ref{secRSP} studied the persistence of one crop variety in a farmers' network of seed exchange.

\section{Model}
\label{secModel}
	\subsection{Notations} 
	
The following notations are used:\\

\begin{tabular}{ll}
 \hline
 $n$& number of patches (farms) \\
 \hline
 $G$& interaction network between patches \\
 \hline
 $p$& density of the network\\
\hline
$n_{edges}$& number of edges\\
 \hline
 $e$& extinction rate\\
 \hline
 $c$& colonisation rate\\
\hline
$n_{gen}$ & number of studied generations\\
\hline
$Z_t$ & state of the system \\
\hline
$\#Z_t$& number of occupied patches for state $Z_t$\\
\hline
$\P(\#Z_{t}>0)$) & Probability of persistence at generations $t$\\ %reellement necessaire ?
\hline
$\E(\#Z_{t})$ & Expected number of occupied patches at generations $t$\\ %idem ?
\hline
 \end{tabular}

\input{Model.tex}

\subsection{Methods for simulations}
\label{subsecMethodsimu}

Since the model is a Markov chain in a finite state space, simulating is quite easy. Hence, the probability of persistence after $100$ generations $\P(T_0> 100)$
and the mean number of occupied patches at the $100^{{th}}$ generation $\E(\#Z_{100})$
can be estimated.
However, in cases where persistence is very likely or very unlikely, a large number of simulations are necessary
to achieve precision in the estimate of the persistence probability. Indeed, we can face two kinds of rare events: rare extinction, or rare persistence.
Some techniques related to the estimation of probabilities of rare events were used. They are based on importance sampling and interacting particle systems.

\subsubsection{Rare persistence}

A very simple interacting particle system \citep{delmoraldoucet2009} is efficient in this case. The idea is to consider simultaneous trajectories (particles) and regenerate
the ones which have been trapped in the coffin state (extinction) among the surviving particles.

\medskip

\noindent \textbf{Algorithm 1}
\begin{itemize}
 \item \textbf{Initialisation:} $N$ particles set at $Z_0^i=(1,\ldots,1)$ for any $i=1,\ldots, N$.
 \item \textbf{Iterations:} $t=1,\ldots,100$:
 \begin{itemize}
 \item \textit{Mutation} Each particle evolves independently according to the Markov model (obtaining $\tilde Z_t^i$ from $Z_{t-1}^i$
 by simulation).
 \item \textit{Selection/Regeneration:} If $\tilde Z_t^i=0$, then $Z_t^i$ is randomly chosen among the surviving particles 
 $\tilde Z_t^j\not=0$. Otherwise $Z_t^i=\tilde Z_t^i$.\\
 Compute $\#E_t=\sum_{i=1}^N \mathbb{I}(\tilde Z_t^i=0) /N$.
 \end{itemize}
\end{itemize}

\medskip
Note that the product $\prod_{t=1}^{100}\#E_t$ is then an unbiased estimator of 
$\P(T_0\le 100)$ \citep{delmoraldoucet2009}.
\\
A sufficient number of particles $N$ must be chosen to ensure that not all the particles die during a mutation step.

\subsubsection{Rare extinction}

If the probability of persistence is high, a lot of simulations are necessary to observe at least one extinction. If no extinction
is observed, then the estimate of the extinction probability is zero. 
To improve the estimator, we have to make extinction more likely in the simulation and to apply a correction in the final estimator 
such that the estimator is still unbiased.
Two methods are proposed to achieve this goal: an importance sampling method and a splitting method \citep{rubinotuffin2009}.
\\

The importance sampling is applied on the extinction phase by increasing the extinction rate $e$ in the simulations.
Indeed, since extinction occurs independently in patches, the ratios due to the change in distribution is tractable.
\\

\medskip

\noindent \textbf{Algorithm 2}
\begin{itemize}
 \item \textbf{Initialisation:} $Z_0=(1,\ldots,1)$, a vector $(e_1^{IS},\ldots,e^{IS}_{100})$ with size the number of generations of
 twisted extinction rate (the twisted rate is not necessarily the same throughout generations) is chosen.
 \item \textbf{Iterations:} $t=1,\ldots,100$:
 \begin{itemize}
 \item \textit{Extinction} Extinction is simulated with the corresponding twisted extinction rate $e_t^{IS}$ and the
 ratio is computed as
 \begin{equation}
  \nonumber
 r_t=\left(\frac{e}{e_t^{IS}}\right)^{d_t}\cdot \left(\frac{1-e}{1-e_t^{IS}}\right)^{\#Z_{t-1}-d_t}\,,
 \end{equation}
 where $d_t$ is the number of extinction events which occur at generation $t$ and $\#Z_{t-1}-d_t$ gives the number
 of occupied patches which do not become extinct at generation $t$.
 \item \textit{Colonisation:} Colonisation is applied according to the model.
 \end{itemize}
\end{itemize}

Hence, the unbiased estimator of $\P(T_0\le 100)$ for $N$ simulations obtained according to the previous algorithm ($(Z_t^i)_{t\in\{0,100\}}$
with ratios ($r_t^i)_{t\in\{0,100\}}$, $i=1,\ldots,N$ ) is 
\begin{equation}
\nonumber
\frac{1}{N}\sum_{i=1}^N\prod_{t=1}^{100}r^i_t \times \mathbb{I}( Z_t^i=0)\,.  
\end{equation}
Since the simulations / particles do not interact, the computation can be done in parallel.
Although the variance of this estimator is not tractable in a closed form, it can still be shown 
that the variance is smaller if
the vector $(e_1^{IS},\ldots,e^{IS}_{100})$ is chosen such that $e_t^{IS}$ increases with $t$.
\\

\medskip

Another solution is to use a splitting technique. The rare event, which is extinction here, is split into intermediate less rare events. The extinction corresponds to 
zero occupied patches at generation $100$.
An intermediate rare event is the number of occupied patches being less than a given threshold $S$ at any generation between $0$ and $100$.
A sequence of thresholds $S_1\ge S_2\cdots \ge S_p$ is fixed and the probability of extinction in $100$ generations reads as
\begin{eqnarray*}
  \P(Z_{100}=0)&=&\P(\exists t, \  Z_t=0)\\
  & = &\P(\exists t,\ Z_t\le S_1) \times \P(\exists t,\ Z_t\le S_2|\exists t,\ Z_t\le S_1 ) \\
  & &  \times \dots \times \P(\exists t,\ Z_t=0| \exists t,\ Z_t\le S_p )\,,
\end{eqnarray*}
where $\exists t$ means implicitly $\exists t\le 100$ and $Z_t\le S_p$ means $\#Z_t\le S_p$. 
\\
The algorithm will keep the trajectories that have crossed the first threshold (the trajectories for which there is 
at least one state with a number of occupied patches below $S_1$).  
From these successful trajectories, offspring are generated from the time of the first crossing and then are kept if they cross
the second threshold and so on.
The ratio of the successful trajectories over the total number of simulated trajectories
between threshold $S_{m-1}$ and $S_m$ is used to estimate the probabilities $\P(\exists t,\ Z_t\le S_m|\exists t,\ Z_t\le S_{m-1} )$.
The splitting algorithm we use is in a fixed success setting that is to say the algorithm waits for a given number of regenerated 
trajectories to cross each threshold before moving to the next threshold. Hence, this setting prevents degeneracy of the trajectories 
(no trajectory manages to cross a threshold) and the precision is controlled in spite of the computational effort \citep{amrein2011}.
\\

\medskip

\noindent \textbf{Algorithm 3}
\begin{itemize}
 \item \textbf{Initialisation:} $N$ particles set to $Z_0^i=(1,\ldots,1)$ for any $i=1,\ldots, N$. Choose 
 the sequence of decreasing thresholds $S_1,\ldots S_p$ and the number of successes $n_{success}$. 
 By convention, $S_{p+1}=0$.
 Set the beginning level of trajectories 
 $L_0^i=0$ and starting state $Z_0^i=(1,\ldots,1)$ for $i=1,\ldots,n_{success}$.

 \item For each threshold $S_m$, $1\le m\le p+1$ , set $s=0$ and $k^m=0$ and  
 repeat until $s=n_{succes}$: 
 \begin{itemize}
 \item Do $k^m=k^m+1$.
 \item Choose uniformly $i\in \{1,\ldots,n_{success}\}.$
\item Simulate a trajectory from generation $L_{m-1}^i$ at state $Z_{m-1}^i$: $(Z_t)_{L_{m-1}^i\le t \le 100}$.
\item If there exists $t$ such that $Z_t\le S_m$,  
do
\begin{enumerate}
 \item $s=s+1$,
 \item $L_m^{s}=\inf\{t,\ Z_t\le S_m\}$,
 \item $Z_m^{s}=Z_{L_m^{s}}$.
\end{enumerate}
 \end{itemize}
\end{itemize}

The unbiased estimator of the extinction probability is then:
\begin{equation}
 \nonumber
\prod_{m=1}^{p+1}\frac{n_{succes}-1}{k^m-1}\,.
 \end{equation}
The fixed success setting ensures the non-degeneracy of the trajectories. However, there is no control on the complexity of the algorithm. 
As a by-product, this algorithm also provides estimations of the probabilities that the trajectories cross the intermediate thresholds.
\\

The case of rare extinction is more difficult since there is no obvious method for efficiently computing the probability unlike the case of rare persistence.
In the two algorithms presented above, the efficiency relies on the tuning of parameters, namely the twisted extinction rates in Algorithm 2 and the sequence of thresholds in Algorithm 3. 
In the present study, these parameters have been set manually; the definition of a general tuning strategy is out of the scope of this article.

\section{Network topology}
\label{secNetwork}

In the following we assume that the topology of a network accounts for a kind of social organisation among patches. The main features of a topology are
emphasised in order to make the differences appear clearly.
The topologies we compare are well known in the literature, but we adapt the simulation models in order
to limit the variability by controlling the number of edges. 
Once a number of edges is set (denoted by $n_{edges}$), a topology consists in a way to distribute edges.
To describe the topology of a network, the distribution of the degrees of nodes is pertinent.
We always work under the constraint of a network with a single component. The package \texttt{igraph} \citep{igraph} in \texttt{R} was used for simulating 
networks for
certain topologies and for plotting. We recall that we assume that the network $G$ is non-oriented.

 \begin{figure}
  \begin{center}
\includegraphics[scale=.2]{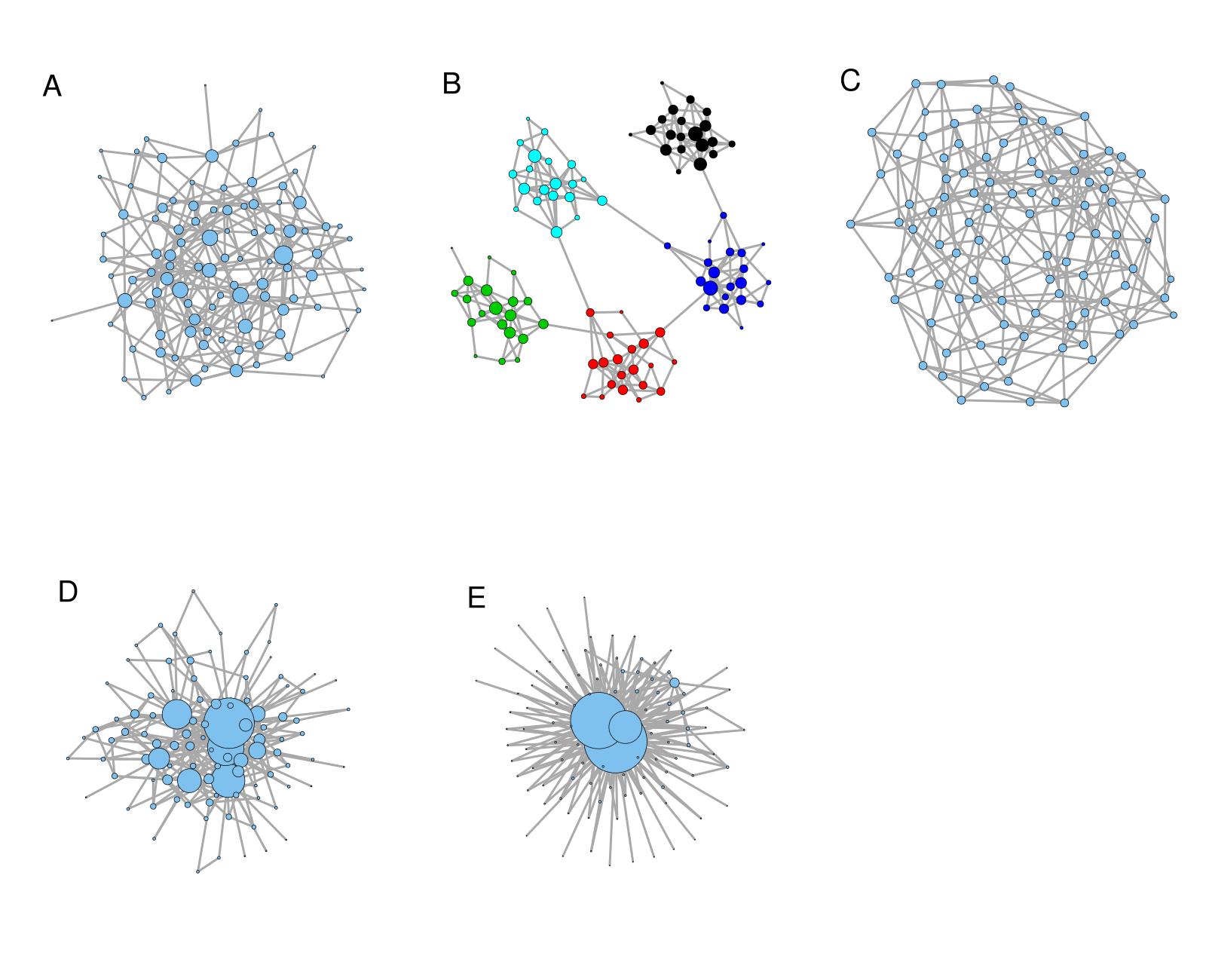}
 \end{center}
\caption{Simulation of networks with $100$ nodes and $247$ edges according to Erd\H{o}s-Rényi model (\textbf{A}), community model (\textbf{B}),
lattice model (\textbf{C}), preferential attachment model with power 1 (\textbf{D}) and power 3 (\textbf{E}). The size of a node is 
proportional to its degree.}
 \end{figure}

\subsection{Erd\H{o}s-Rényi model}

This random graph model was introduced by \citet{erdosrenyi1959} and is defined as follows: for a chosen number of edges $n_{edges}$, the network is constructed by choosing uniformly among all the possible edges $\left(\begin{array}{c}n\\ 2
\end{array}\right)$.
                                   
When the number of nodes is large, the distribution of the degrees of nodes is close to a Poisson distribution \citep{Albert2002}.

\subsection{Community model}

The community model was used to take into account cases where networks are organised through communities. Inside a community, the nodes
are connected with a high probability whereas the connection probability is weak between two nodes belonging to two different
communities. The spirit of this model is drawn from Stochastic Block Model \citep{nowickiSnijders2001}. The community sizes
are set to be equal, the intra-community and the inter-community connection probabilities are the same 
in order to reduce the number of parameters for defining such a network. This model is then tuned
by the number of edges, the number of communities and the ratio of the intra connection probability over the inter connection probability
(this ratio is greater than 1 in order to favour intra connection).

\subsection{Lattice model}

We made an abusive use of the word lattice, by considering graphs with quasi-equal degrees. 
The lattice model stands for organizations built on the basis of the spatial neighborhood.
We propose a simulation model which is flexible since it works for any number of nodes and edges.
The main idea is to fix the smallest upper bound on the degree of a node for given numbers of nodes and edges.
This upper bound is computed as the floor integer number of $2 n_{edges}/n$.
First, a one dimension lattice is created in order to ensure a single component graph. 
Then, edges are added sequentially with a uniform distribution between nodes which have not reached 
the bound on their degree.
Hence, in such a graph, all nodes have quite the same role and importance.

\subsection{Preferential attachment model}

This version of the preferential attachment model was proposed by \citet{barabasi1999}.
It was designed to model growing networks and to capture
the power-law tail of the degree distribution which was noticed in real networks
in many fields of application.
The nodes are added sequentially. In each step, a single node is added and is connected to the nodes already in the network
with probability 
\begin{equation}
 \nonumber
 \mathbb{P}(\textrm{connection to node } N_k)\propto \ degree(N_k)^b\,,
\end{equation}
where the power $b$ is chosen in order to tune the
strength of the preferential attachment. This generative model tends to create nodes with a high degree which
have a central role in the network. It is clearly opposed to the lattice model which 
makes the degrees quasi homogeneous.
\\

To be able to fix the number of nodes and the number of edges independently, we draw uniformly a sequence of edges added
for each new node with the constraint that there is at least one edge (ensuring the single component network) and that
the number of edges to be added is less than the number of nodes already in the network at a given step.

\section{Global analysis of the impact of the topology}
\label{secAnalysisSensitivity}

\input{analysesensi.tex}
\section{Application to seed diffusion among farmers: the case study of the emergence of the Réseau Semences Paysannes}

\label{secRSP}

\subsection{Context}
\begin{figure}[ht]
		    \subfloat[][]{\label{sen}\includegraphics[scale=0.25]{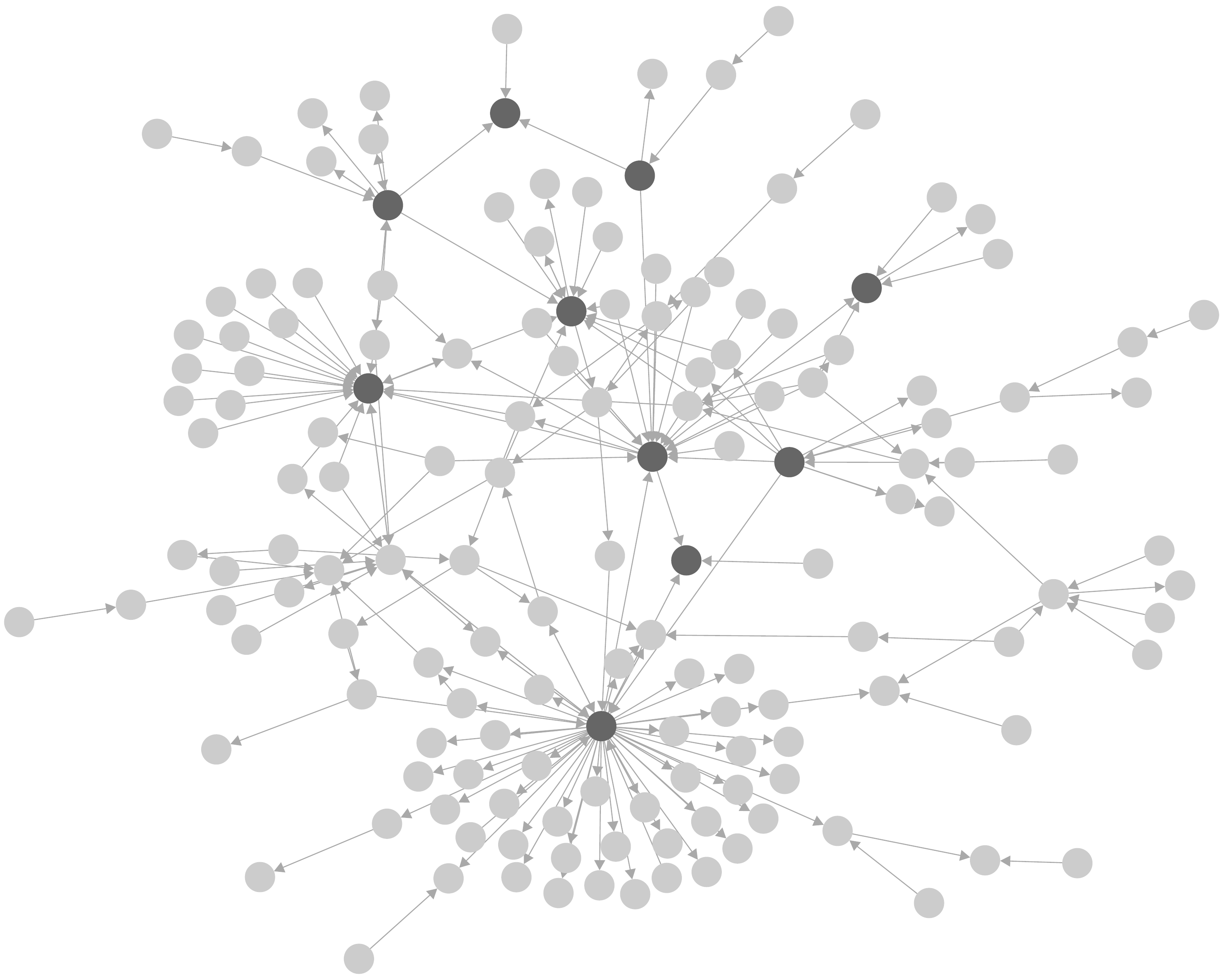}}
		    \subfloat[][]{\label{sen2}\includegraphics[scale=0.25]{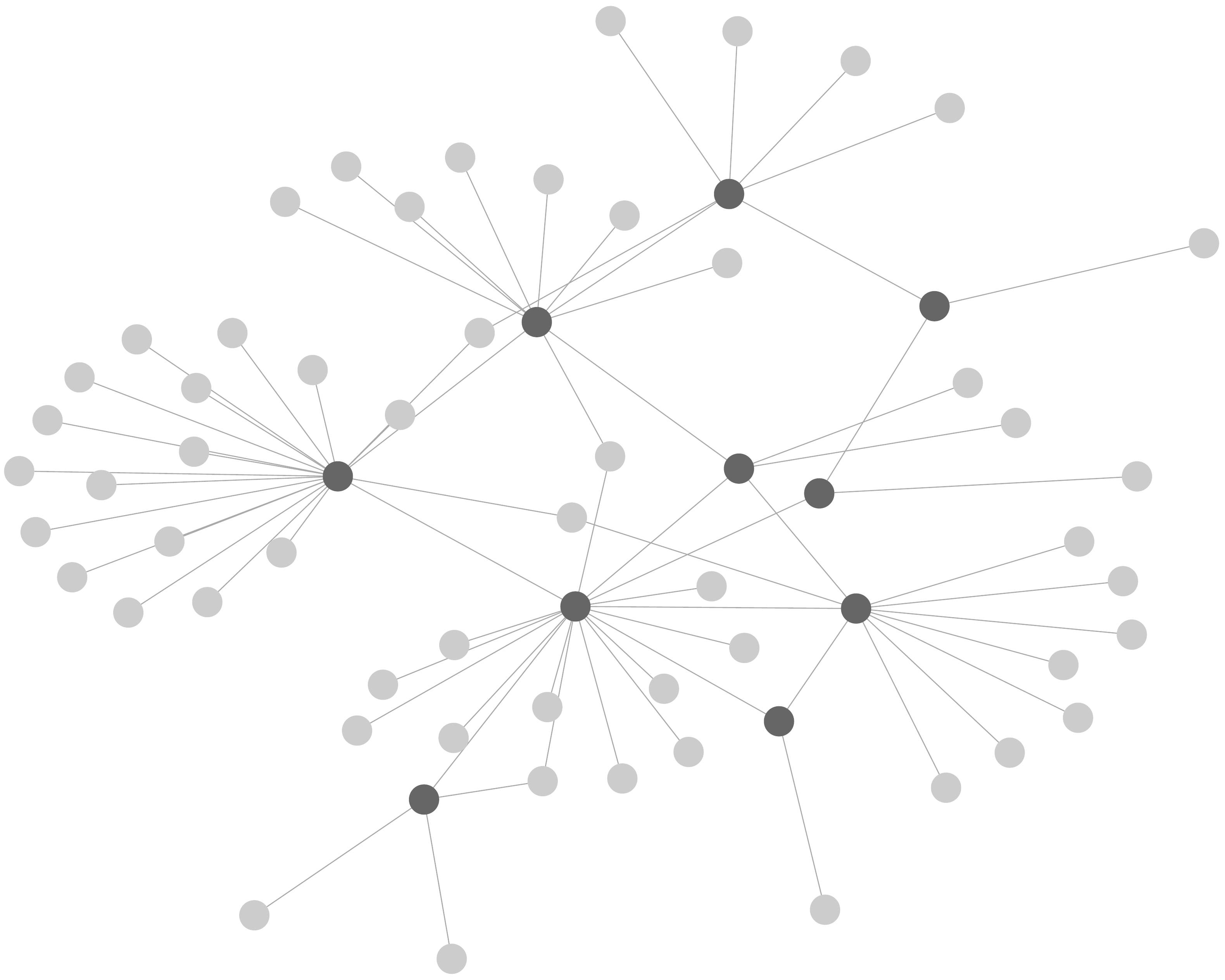}}
	\caption{\protect\subref{sen} Summary network of bread wheat seed circulation among 152 farmers drawn from data collected based on 10 interviews covering a period from 1970 to 2005.
	\protect\subref{sen2} Subgraph of the reliable seed circulation events from 1970 to 2005 based on the 10 interviews and used to estimate $\hat{p}_{50}$.  Interviewed people are in dark grey and mentioned people in light grey.}
\end{figure}

Our first application of the sEC model was to describe an emergent farmers' movement involved in seed exchange of crops and vegetables in France.
From the beginning of the 1990's in Europe, new farmers' organisations have emerged with the aim of sharing practices and seeds \citep{bocci_peasant_2008}.
In a preliminary study, Demeulenaere and Bonneuil identified the global social dynamics in the context of the ``Réseau Semences Paysannes'' (RSP), a French national farmers' organisation created in 2003 \citep{demeulenaere_semences_2012}.
They described this social movement highlighting emergent rules and giving a semi-quantitative picture of the dynamics of this social organisation. 
They focused their study on one of the RSP's subgroups specialized in bread wheat (\textit{Triticim aestivum}). 
Based on informants of the RSP, they identified key actors. They performed 10 exhaustive interviews to collect data on which varieties were present in the fields of the farmers
and from whom they were obtained. Additional information such as to whom farmers provided varieties was less informed.
They completed data collection with 8 additional semi-directive interviews and 7 phone interviews.    
They collected 778 distinct records of seed exchange events among 160 actors between 1970 and 2005.
These seed exchanges involved 175 different varieties of bread wheat.  
After pooling all the information collected between 1970 and 2005, a directed seed circulation network was drawn where an edge connects two farmers 
who have carried out at least one seed diffusion event during this period.
Three connected components were identified: one giant component (152 nodes, Fig. \ref{sen}) and two small ones (5 and 3 nodes respectively, not shown).
The average colonisation rate ($c$) was estimated as the number of diffusion events per variety per farmer per year. 
The number ranged over time from 0.03 to 0.66. 

\subsection{Question, approach and assumption}
In the context of an emergent self-organized system, a crucial question is to what extent do changes in social organisation impact the global ability of the system to maintain varieties?

Relying on our knowledge about RSP evolution, three network topologies and two network sizes were simulated to represent evolution of this social organisation, assuming that seed exchange networks were embedded in more complex social networks.
Five scenarios were defined to provide a framework for studying the impact of such social change. They are described in the next section. 
Then, the sEC model was used to represent the dynamic process of seed circulation and extinction for each network topology.
A dedicated sensitivity analysis was designed to explore specific ranges of $e$ and $c$ in a short time window. 
Working at this time scale was motivated by the rapid change in social organisation of such systems. 
The probability of persistence and the expected number of occupied patches was assessed for each scenario to compare the ability of the system to maintain the variety circulating in the network.

Using the sEC model in the context of seed systems assumed that farmers always wanted to recover the variety after losing it. 
In addition, we assumed that all farmers had the same behaviour, having the same ability: 1) to host a seed lot of a variety through the seed diffusion process
(uniform colonisation probability, $c$) and 2) to lose it through a stochastic process of extinction (uniform extinction probability, $e$). 
This assumption was made to highlight the position in the network independently of individual characteristics.

\subsection{Scenarios of evolution}

Rapid evolution of the social organisation can be qualitatively depicted through three main phases. This description relies on a participant observation study \citep{demeulenaere_semences_2012} during farmers' meetings between 2003 and 2012.
During the first phase, several dozen farmers were invited to participate in national meetings. 
At the beginning, only a few exchanged seed with a limited knowledge of each other. 
For this reason, we assumed random seed exchanges among farmers using an Erd\H{o}s-Renyi network (ER). 
After several meetings, a few farmers became more popular and more central in seed diffusion. 
We modelled this stage using a preferential attachment (PA) algorithm for designing the network topology and accounting for a more important role of a few farmers.
Eventually, the number of farmers exchanging seeds highly increased. At the same time, a change was observed from national meetings to more local events thanks to the creation of local associations involved in seed exchange (from 0 to 17 between 2003 and 2012). 
We considered the community model (COM) following the stochastic block model as an appropriate network model to mimic this new organisation with most seed exchange at the local scale (within groups) and rare events at the global scale (long distance and among groups).
Based on these observations, five scenarios were defined for analysing the impact of change in social organisation on the ability of the self-organised system to maintain varieties (Table \ref{scenarios}):
\begin{itemize}
	\item 1: random seed exchanges among few farmers (ER:50)
	\item 2: scale-free seed exchanges among few farmers (PA:50)
	\item 3: community-based seed exchanges among many farmers (COM:500)
	\item 4: random seed exchanges among many farmers (ER:500)
	\item 5: scale-free seed exchanges among many farmers (PA:500)
\end{itemize}
First, we compared the results obtained for scenarios 1 and 2 to study the evolution of the system capacity to maintain varieties after a change in social organisation in the context of small size population.
Then, scenario 3 was compared to scenarios 4 and 5 to understand the consequence of a new social configuration in maintaining crop diversity after an increase in the network size. 

\subsection{Sensitivity analysis}

This additional sensitivity analysis is required to draw conclusions about the scenarios in the context of the RSP study (different 
number of patches, ranges of $e$ and $c$ and a shorter time horizon). 
To initialize the simulations, each actor owned only one and the same variety. We defined three levels of event frequency to mimic different global behaviours in terms of seed circulation and maintenance: low frequency with $e=0.1$, intermediate frequency with $e=0.5$ and high frequency with $e=0.8$.
It was chosen to investigate two variety statuses: popular and rare varieties. 
We considered that rare varieties were less diffused with an $e/c$ ratio of 5 compared to 1 for the popular ones.
We fitted the density of the small network ($n=50$) to the density of the observed network (\ref{sen2}) which gives $\hat{p}_{50}=0.21$

For scenarios $3-5$ with networks of size 500, density was considered equal to $\hat{p}_{500}=\hat{p}_{50}/(500/50)=0.021$, considering that people shared the same average degree whatever the size of the network.

This framework allowed us to investigate the impact of topological properties of relational networks on the dynamic of the system and more specifically on the probability of persistence of varieties after 30 generations, $\P(\#Z_{30}>0)$, and the relative expected number of occupied patches after 30 generations, $\E(\#Z_{30})$. The choice of 30 generations corresponded to the time scale of observed seed exchange. In addition, we considered that such social organisation in the context of emergent social movements is rapidly evolving without reaching a real equilibrium. 
Longer simulations seemed to be less informative for understanding properties of this self-organised system. 

	\begin{table}[h]
	  \begin{small}
	    \begin{center}
	      \begin{tabular}{ccccccc}
		  \hline
	  Scenario 	& Comparison	&	$n_{vertex}$ & $n_{edge}$	& 	topo 		&     $e$	& ratio $e/c$	\\ 
		  \hline{}
	  1a 		& 	1	&	50	&	263		&	ER	& $\{0.1;0.5;0.8\}$	& 1	\\  
	  1b 		& 	1	&	50	&	263		&	ER	& $\{0.1;0.5;0.8\}$	& 5	\\  
	  2a 		& 	1	&	50	& 	263		&	PA	& $\{0.1;0.5;0.8\}$ 	& 1	\\  
	  2b 		& 	1	&	50	& 	263		&	PA	& $\{0.1;0.5;0.8\}$ 	& 5	\\  
	  3a 		& 	2	&	500	& 	2682		&	COM$^*$	& $\{0.1;0.5;0.8\}$	& 1	\\
	  3b 		& 	2	&	500	& 	2682		&	COM$^*$	& $\{0.1;0.5;0.8\}$	& 5	\\
	  4a 		& 	2	&	500	& 	2682		&	ER	& $\{0.1;0.5;0.8\}$	& 1	\\
	  4b 		& 	2	&	500	& 	2682		&	ER	& $\{0.1;0.5;0.8\}$	& 5	\\
	  5a 		& 	2	&	500	& 	2682		&	PA	& $\{0.1;0.5;0.8\}$	& 1	\\
	  5b 		& 	2	&	500	& 	2682		&	PA	& $\{0.1;0.5;0.8\}$	& 5	\\
		  \hline
	      \end{tabular}
	    \end{center}
	  \end{small}
	  \caption{Description of the 5 scenarios ($^*$:the COM model is defined for 10 groups of 50 farmers with a probability of connecting people from the same community 10 times higher than
	  the probability of connecting two people from different communities).}

	  \label{scenarios}
	\end{table}

\subsection{Results}

	\begin{table}[h]
	  \begin{small}
	    \begin{center}
	      \begin{tabular}{cccc}
		  \hline
	  	&  $e$	&	$\P(\#Z_{30}>0)$ 	&$\E(\#Z_{30})$ 	\\		
		  \hline{}                                          
$e/c=1$		&0.1	&	$ER = PA = 1$	&$ER \sim PA=44$	\\
		&0.5	&	$ER = PA = 1$		&$ER \gtrsim PA=44$	\\
		&0.8	&	$ER=0.9 > PA= 0.7$		&$ER=37 > PA=25$		\\
$e/c=5$		&0.1	&	$ER = PA=1$		&$PA \gtrsim ER=25$	\\
		&0.5	&	$PA=0.8 \gg ER=0.3$		&$PA=13\gg ER=3$		\\
		&0.8	&	$PA = ER=0$			&$PA=ER=0$			\\
		  \hline
	      \end{tabular}
	    \end{center}
	  \end{small}
	  \caption{Summary results of persistence probability ($\P(\#Z_{30}>0)$) and expected number of occupied farms after 30 generations ($\E(\#Z_{30})$) for 50 farmers, with $\sim$: difference lower than $2\%$, $>$: difference between $10-50\%$, $\gg$: difference higher than $50\%$.}
	  \label{n50}
	\end{table}

	\begin{table}[h]
	  \begin{small}
	    \begin{center}
	      \begin{tabular}{cccc}
		  \hline
	  	& $e$		& $\P(\#Z_{30}>0)$ 	&$\E(\#Z_{30})$ \\ 
		  \hline{}  
	$e/c=1$	&	0.1	&$PA= ER = COM=1$		&$ER \sim COM \gtrsim PA=425$\\
	       &	0.5	&$PA= ER = COM=1$		&$ER \sim COM \gtrsim PA=427$\\
	       &	0.8	&$PA \sim ER = COM=1$	&$ER \sim COM=382 >PA=314$\\
	$e/c=5$&	0.1	&$PA= ER = COM=1$		&$ER \sim COM \sim PA=249$\\
	       &	0.5	&$ER\sim COM \sim  PA=1$	&$PA=193 \gg ER \gg COM=40$\\
	       &	0.8	&$PA=0.5 \gg ER= COM=0$		&$PA=43>ER=COM=0$\\
		  \hline
	      \end{tabular}
	    \end{center}
	  \end{small}
	  \caption{Summary results of persistence probability ($\P(\#Z_{30}>0)$) and expected number of occupied farms after 30 generations ($\E(\#Z_{30})$) for 500 farmers, with $\sim$: difference lower than $2\%$,$\gtrsim$: difference between $2-10\%$, $>$: difference between $10-50\%$, $\gg$: difference higher than $50\%$.}
	  \label{n500}
	\end{table}

\paragraph{Small networks: scenarios 1 and 2} 
The change in network topology was the main difference between scenarios 1 and 2.
For the popular varieties, ($e/c=1$), PA only showed a lower probability of persisting ($\P(\#Z_{30}>0)$) and a lower expected number of occupied farms ( $\E(\#Z_{30})$) for the higher values of $e$ and $c$ compared to ER (Table  \ref{n50}).
For rare varieties, more susceptible to extinction ($e/c=5$), an inversion between ER and PA was observed in terms of ability to maintain the resource ($\P(\#Z_{30}>0)$) for intermediate values of $e$ and $c$, before decreasing to zero for the highest values (Table  \ref{n50}). 
This trend was confirmed by the expected number of occupied patches.

In both cases ($e/c=1$ and $e/c=5$), the results are consistent with section \ref{secAnalysisSensitivity}:
the balanced distribution of degree in ER networks conferred a higher persistence probability and a higher relative occupancy compared 
with a more hierarchical organisation (PA) in the context of safe situations. It is the heterogeneity of the degree distribution 
that conferred more persistence in the context of critical extinction.

\paragraph{Larger networks: scenarios 3, 4 and 5} They are characterized by the larger number of actors.
COM configuration was compared to initial topologies: ER and PA networks.
Simulation results showed an equivalent $\P(\#Z_{30}>0)=1$, whatever the frequency of event (low or high $c$ and $e$) and the network topology (Table \ref{n500}) for popular varieties ($e/c=1$).
Thus, with such parameter values it was not likely that a variety disappeared whatever the topology.
Increasing the number of actors from 50 to 500 induced a substantially higher expected number of occupied farms for ER and COM topologies compared to PA (Table \ref{n500}). The opposite behavior was observed for rare varieties ($e/c=5$).
We noticed that ER and COM provided similar results whatever the conditions. 

\subsection{Role of the social network topology on variety persistence and recommendations}
\paragraph{Role of the social network topology} 
Different factors influenced the distribution and persistence of varieties.
The number of farmers as well as $e$ and $c$ parameters were obviously the most important ones.
The increase in the number of participating farmers from the creation of the RSP 
has substantially improved the probability of maintaining rare and popular varieties within the system. 
The network topology did not always have an incidence on persistence. 
When it was the case, it was not always the same topology that outperformed the others depending on the situation.
Such behaviour depended on the status of the variety under consideration. Popular varieties were better maintained with ER or COM topologies
because of the balanced degree that avoids local extinctions, whereas rare varieties persisted better with PA topology. In the case of rare varieties,
PA topology with few farmers as hubs allowed the variety to be quickly redistributed through the network after local extinctions.  

Relying on simulation results, we showed that the self-organized trajectory of the RSP from small ER, then to small PA to large COM improved the efficiency
of the system at maintaining popular varieties compared with rare varieties.
The COM network model seems to be a realistic topology for large networks since local meetings with a subset of the farmers are easier to organize.
In this context, a community is driven by the local meetings and farmers participating in the same meeting are likely to be connected.
In COM, very few farmers are linked to farmers from other communities. 
Nevertheless, we observed that it led to the same ability for persistence and occupancy as did the ER network. 
These findings illustrated the independence between the ability to maintain a variety and the connection within community and across communities in the context of popular varieties. 
A detailed sensitivity analysis of the COM parameters would allow one to assess whether some COM configurations depart from ER behaviour. 
This sensitivity analysis could be extended to the study of a mixture model with COM and PA topologies. 
Such topologies would allow one to model an even more realistic situation accounting for a persistently higher degree of a few farmers within and across communities. 

It was not possible to forecast the behaviour of the model using only the $e/c$ ratio due to complex interactions with the size of the network and the type of topology. 

\paragraph{Recommendations for future studies on seed systems}
Sensitivity analysis on the extinction-colonisation model confirmed that $e$ and $c$ parameters were the most contributing factors to the ability of the system to maintain a variety. 
Topological parameters like the density of the network also looked important. Unfortunately, such data have not yet been collected. 
One of the reasons is that seed systems are rapidly changing, often informal or even illegal depending on the country legislation. 
Nevertheless, our work showed that a good knowledge of event frequency (extinction rate and seed exchange rate), of the status of the variety (rare or popular) in addition to the density of the social network could provide important clues to the health of the seed system. 
Thus, particular attention has to be paid to these particular quantities in future studies to strengthen our understanding of the sustainability and health of seed systems.

\section{Conclusion}

This study aimed to investigate the role of the network topology in a dynamic extinction colonisation model. 
Obviously, the number of edges was the most important feature of the network. 
The topology (distribution of the edges) impact was less important but not negligible and its impact depended on the other parameters (extinction rate $e$, colonisation rate $c$ and number of edges).
In section \ref{secModel}, we have highlighted the limits of describing a stochastic dynamic extinction colonisation model only by the ratio $e/c$. 
As noticed in section \ref{secAnalysisSensitivity}, if the relevant parameters led to a probable extinction, networks with high degree nodes (PA) were more resistant than networks with balanced degrees (LAT, ER or COM). 
On the contrary, if persistence was quite certain, more patches were occupied in balanced networks than in the PA networks.
The community structure (COM) and ER showed similar properties with respect to persistence and occupancy.
These results obtained for small networks and after a short time period were consistent with those obtained for large networks after reaching a quasi-equilibrium state as shown by \cite{Gilarranz201211}.
Nevertheless, the necessity to properly estimate the persistence probability and the expected occupancy in strongly stochastic conditions was pointed out and a specific procedure was provided.
\cite{Franc2004,Peyrard2008} proposed more accurate approximations of the sEC model to determine the behaviour of the system close to critical
situations. They demonstrated the importance of particular geometrical features of the network such as the clustering coefficient and 
the square clustering coefficient for describing the impact of the network in the evolution of the system.
Further studies should be conducted to determine the role of these features in a limited network size.

Such work could contribute to feeding the thoughts for further discussions with farmer organisations and community seed systems, particularly on the way to monitor popular and rare varieties circulating in the system. 
This study improved our understanding of the role of the social organisation in maintaining crop diversity in such emergent self-organised systems.

\section{Acknowledgements}

The authors of the paper want to thank the farmers of the Réseau Semences Paysannes who accepted to devote a part of their time discussing the questions of seed exchange.
The authors also thank the participants of the NetSeed and MIRES consortia for their fruitful discussions. NetSeed is a project funded by the Fondation pour la Recherche sur la Biodiversité (FRB), Paris, France and hosted by the Centre de Synthèse et d'Analyse sur la Biodiversité (CESAB), Aix-en-Provence, France. MIRES is a working group on modelling seed exchanges funded by the Réseau National des Systèmes Complexes, Paris, France. Mathieu Thomas is a postdoctoral fellow who obtained the grant ``Contrat Jeunes Scientiques''
from the Institut National de la Recherche Agronomique, France.

\newpage

\pagebreak

\bibliographystyle{apalike}
%\bibliography{refExtinction}

\end{document}

%% file: Model.tex
%%%%%%%%%%%%%%%%%%%%%%%%%%%%%%%%%%%%%%%%%%%%%%%%%%%%%%%%%%%%%%%%%%%%%%%%
\subsection{Model definition}
%\subsection{Notations}
%%%%%%%%%%%%%%%%%%%%%%%%%%%%%%%%%%%%%%%%%%%%%%%%%%%%%%%%%%%%%%%%%%%%%%%%

%\SR{node = 'population' = 'farm' = 'patch': could we reduce the number of different names?}

The model describes the presence or absence of a crop variety
on $n$ different farms (patches according to metapopulation
vocabulary) during a discrete time evolution process. 
%We consider a metapopulation made of $n$ patchs, each corresponding to a farm. 
This metapopulation is identified with a network $G$ with $n$ nodes (farms) and adjacency matrix $A = [a_{ij}]_{i, j}$ were $a_{ij} = 1$ if patches $i$ and $j$ are connected ($i \sim j$) and $0$ otherwise. 
% +The adjacency matrix of this relational network is denoted by $A=(a_{i,j})_{1\le i,j\le n}$ where $a_{i,j}=1$ if the patches $i$ and $j$ are linked, $a_{i,j}=0$ otherwise. Its diagonal elements are equal to $0$ ($a_{i,i}=0$).  
This matrix is symmetric which means that a relation between two patches is reciprocal.  
We further denote by $Z_{i, t}$ the occupancy of patch $i$ ($i = 1 \dots n$) at time $t$, namely $Z_{i, t} = 1$ if patch $i$ is occupied at time $t$ and $0$ otherwise. 
The vector $Z_t = [Z_{i, t}]_{i}$ depicts the composition of the whole metapopulation at time $t$.

A time step corresponds to a generation of culture.  Between two generations, two events can occur: extinction and colonisation with respective rates $e$ and $c$. 
Within each time step, extinction events first take place and occur in occupied patches independently of the others, with a probability $e$, supposed to be constant over patches and time.
Colonisations events then take place and are only possible between patches linked according to the static relational network $G$.
An empty patch can be colonised by an occupied patch with a probability $c$.  This probability is also assumed constant over linked patches and time steps.
Thus, the probability that the patch $i$, if empty at generation $t$, is not colonised between generations $t$ and $t+1$ is equal to $(1-c)^{o_{i,t}}$ 
where $o_{i,t}$ is the number of occupied patches at generation $t$ linked to the patch $i$: $o_{i,t}=\sum_j a_{ij} Z_{j,t}$. 
This model is similar to the one proposed in
\cite{Gilarranz201211} and also to the epidemic model used in
\cite{Chakrabarti2008}.  It can also be seen as a particular case of
the models discussed in \cite{Adler1994,Day1995,Hanski2000}.

% As for the dynamics of the exchange along the network, we denote by $e$ is the extinction rate
% and by $c$ is the colonization rate.

% We use the following notations:
% \begin{itemize}
% \item $n$ is the number of populations constituting a
%                   metapopulation, with a population corresponding to a
%                   farm
% \item $Z_{i,t}$ is the composition of the population
%                   %ou patch $i$ in generation $t$: , where $Z_{i,t}=0$
%                   if variety is not present in the population,
%                   $Z_{i,t}=1$ else
% \item $Z_t$ is the composition of the metapopulation
%                   in generation $t$.
% \item $e$ is the extinction rate
% \item $c$ is the colonization rate
% \item $G=(V,E)$ describes the network composed by a
%                   set of $V$ nodes and a set of $E$ edges ($V=n$)
% \item $A$ is the adjacency matrix of the network $G$
%                   (size $n\times n$)
% \item $M$ is the stochastic matrix of our model (size $2^n\times 2^n$), describing transitions in the Markov chain
% \item% $R$ is obtained by renormalization (to make the
                     % row sum to $1$) of the submatrix of $T$ where
                     % the first row and the first column have been
                     % deleted (size $2^n-1\times 2^n-1$)
% \item $\lambda_{i,B}$ are the $i^{th}$ eigenvalues of matrix $B$
% \end{itemize}

%%%%%%%%%%%%%%%%%%%%%%%%%%%%%%%%%%%%%%%%%%%%%%%%%%%%%%%%%%%%%%%%%%%%%%%%
%\subsection{Model definition}
\subsection{Model properties}

As recalled in \cite{Day1995}, the stochastic process $(Z_t)_{t\in\mathbb{N}}$ is a discrete time
Markov chain with $2^n$ possible states. The matrices describing the colonisation
$C$ and the extinction $E$ can be constructed and the transition
matrix of $(Z_t)_{t\in\mathbb{N}}$ is obtained as the product of these
two matrices: $M=E\cdot C$.  
We assume here that $Z_t$ is irreducible and aperiodic which is ensured if the adjacency matrix
$A$ of the social network has only one connected component.  
In the sequel, we denote by $\lambda_{B, k}$ the $k^{th}$ eigenvalue of any matrix $B$. 
Indeed, the leading eigenvalue of $M$ is $\lambda_{M,1}=1$, its multiplicity is
$1$ and the corresponding eigenvector is the stationary distribution.
If $e>0$, this unique stationary distribution consists of being stuck in one state, 
%in that case, a vector with a $1$ and $0$ elsewhere.  
% This state is
denoted by $0$ and called absorbing state or coffin state. The coffin state corresponds to all patches empty which means that the variety
is extinct. Thus, if $Z_t=0$, for any $s>t$, $Z_s=0$.  We denote by
$T_0$ the extinction time:
\begin{equation}
 \nonumber
T_0=\inf\{t>0,\#Z_t=0 \}\,.
 \end{equation}

Since the number of states is finite, $\P_z(T_0<\infty)=1$ for any
initial state $z$ ($\P_z$ denotes the probability measure associated
with the chain $Z_t$ and initial state $Z_0=z$).  The second
eigenvalue $\lambda_{M,2}$ governs the rate of convergence toward the
absorbing state, i.e.
\begin{equation}
 \label{convAS}
\P_z(T_0>t)=\P_z(\#Z_t>0) =O(\lambda_{M,2}^t)\,.
\end{equation}
The smaller this eigenvalue is, the faster the convergence is. Hence,
we can study the probability of extinction in a given number of
generations or the mean time to extinction for an initial condition on
occupancies at generation $0$, a network and a set of parameters.\\

%%%%%%%%%%%%%%%%%%%%%%%%%%%%%%%%%%%%%%%%%%%%%%%%%%%%%%%%%%%%%%%%%%%%%%%%
\subsubsection{Quasi-stationary phase}
Although extinction is almost sure, the probability of reaching
extinction in a realistic number of generations can still be small. In
that case, we aim to study the behaviour of this dynamic before
extinction.  In some cases, the Markov chain $Z_t$ conditioned to
non-extinction $\{T_0>t\}$ converges toward a so-called
quasi-stationary distribution \citep{Darroch1965,Meleard2012}.  
This
quasi-stationary distribution exists and is unique provided that $Z_t$
is irreducible and aperiodic.  
Note that quasi-stationary distribution may also exist in reducible chains \citep{Vandoorn2009}. 
The transition matrix on the transient states, denoted by $R$, has
dimension $(2^{n}-1)\times(2^{n}-1)$ and is defined as a sub-matrix of
$M$ by deleting its first row and its first column, corresponding to the coffin state. If it exists, the
quasi-stationary distribution is given by normalizing the eigenvector
of the reduced matrix $R$ associated with its leading eigenvalue
$\lambda_{R,1}$. We denote by $\alpha$ this distribution over the
transient states.  It can be noticed that
$\lambda_{R,1}=\lambda_{M,2}$. As stated in
\cite{Darroch1965,Meleard2012},
\begin{equation}
 \label{convQSD}
\sup_{z,z' \textrm{ transient\ states
}}|\P_z(Z_t=z'|T_0>t)-\alpha_{z'}|=O\left(\left(\frac{|\lambda_{R,2}|}{\lambda_{R,1}}\right)^t\right)\,.
\end{equation}
Therefore, the quasi-stationary distribution is met in practice if the
Markov chain converges faster toward it than toward the absorbing
state which corresponds to
${|\lambda_{R,2}|}/{\lambda_{R,1}}<<\lambda_{R,1}$.  \\

Building the transition matrix allows an exact study of the dynamic of
the variety persistence. However, due to its large size: $2^{n}\times
2^{n}$, building such a matrix and seeking its eigenvalues is not
possible for $n>10$.  Therefore, for bigger $n$, we have to run
simulations.

%%%%%%%%%%%%%%%%%%%%%%%%%%%%%%%%%%%%%%%%%%%%%%%%%%%%%%%%%%%%%%%%%%%%%%%%
\subsubsection{Large network approximation}
Another solution is to use an approximate version of the model as
proposed by \cite{Chakrabarti2008}.  They describe the recurrence
relation between the probabilities of occupancies at generation $t+1$
and these probabilities at generation $t$. In the computation of the
recurrence relation, they consider the occupancies of the patches at
generation $t$ as independent of each other. Thus, their relation involves only the
$n$ patches and not all of the $2^n$ possible configurations.  This
approximation leads to the following relation:
\begin{equation}
 \label{relationreccurence}
p_{i,t+1}=1-\zeta_{i,t+1}p_{i,t}e-\zeta_{i,t+1}(1-p_{i,t})\,,
\end{equation}
where $p_{i,t}$ is the probability of occupancy of patch $i$ at
generation $t$ and $\zeta_{i,t}$ is the probability that patch $i$ is
not colonised at generation $t$. The following equation derives from the independence
approximation:
\begin{equation}
 \nonumber
\zeta_{i,t+1}=\prod_{j\sim i}\left(1-cp_{j,t}\right)\,.
 \end{equation}

%where $i\sim j$ means that patches $i$ and $j$ are linked.  
From this
approximation, they derive a frontier between a pure extinction and an
equilibrium phase depending on $e$, $c$ and $\lambda_{A,1}$ the
leading eigenvalue of the adjacency matrix $A$ of the network.  If
$e/c$ is above $\lambda_{A,1}$, a pure extinction shall take place, if
it is below, the patch occupancy shall reach an equilibrium where
the number of occupied patches varies around a constant number.  More
specifically, if ${e}/{c}>\lambda_{A,1}$, the occupancy
probabilities $(p_{i,t})$ tend to $0$ ($0$ is a stable fixed point).
Moreover, in the case where ${e}/{[c(1-e)]}>\lambda_{A,1}$, the
decay over time of the $p_{i,t}$ is exponential,
$p_{i,t}=O((1-e+c(1-e)\lambda_{1,A})^t)$ for any $1\le i\le n$. 
Otherwise, if ${e}/{c}<\lambda_{A,1}$, there exists a fixed point with non-zero probabilities of
occupancies. This non-zero equilibrium clashes with the almost sure convergence of the Markov chain toward the coffin state. 
\\

\begin{figure}
 \begin{center}
 \includegraphics[width=.5\textwidth]{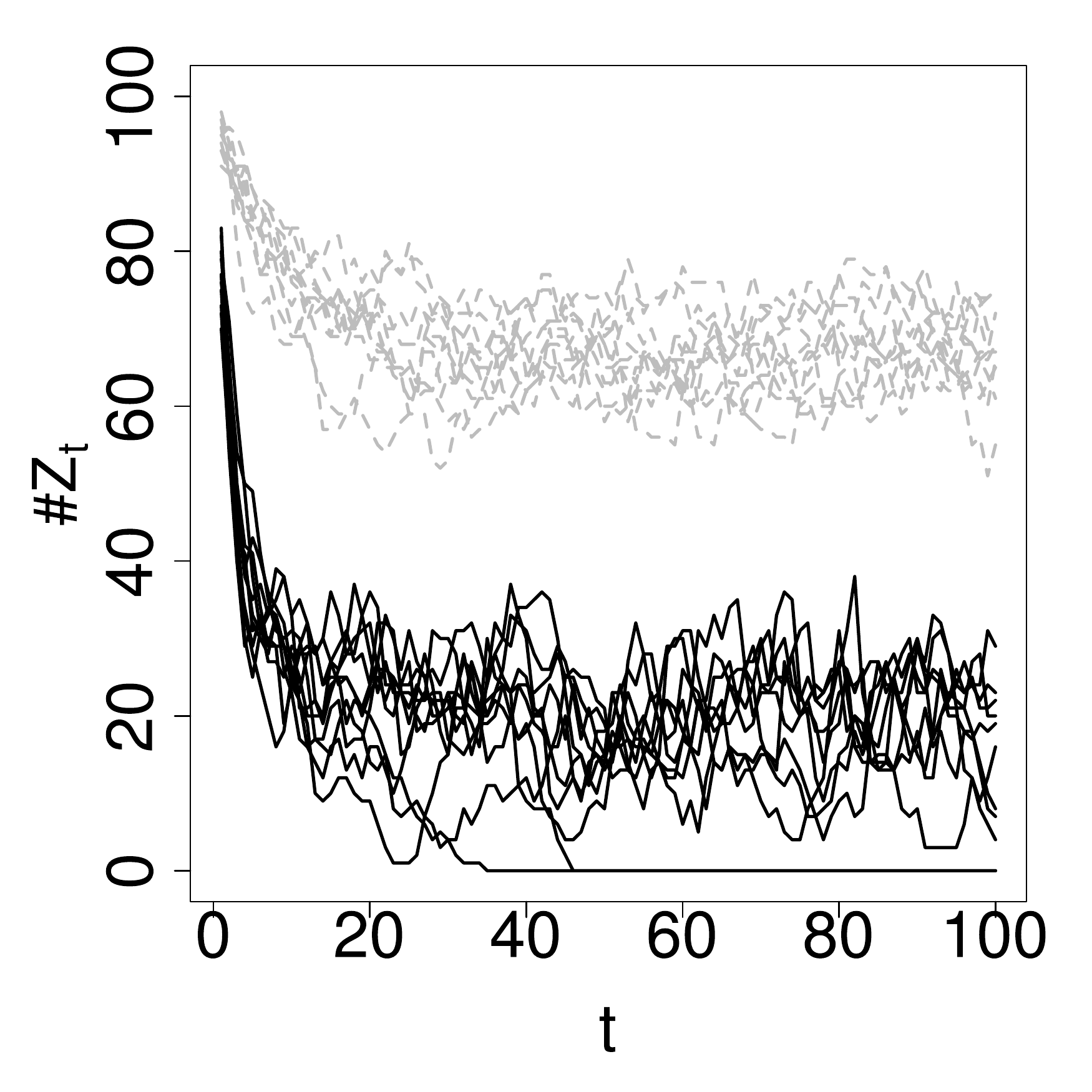}
 \end{center}
\caption{Number of occupied patches for replications from the dynamic model over $100$ generations, network fixed and parameters fixed
at $c=0.05$ and $e=0.25$ (black solid lines) or $e=0.05$ (grey broken lines). The initial state was chosen such that all patches are occupied.}
\label{figevolution}
 \end{figure}

The frontier ${e}/{c}=\lambda_{A,1}$ is also found to be a
relevant threshold for persistence in deterministic models such as
the Levins model \citep{Levins1969} and its spatially realistic
versions \citep{Hanski2000,sole_self-organization_2006}. In a stochastic model,
extinction eventually takes place since there is an absorbing state.
From the previous statements on the quasi-stationary distribution,
observing an equilibrium phase on simulations as in
\cite{Gilarranz201211} actually corresponds to a phase where the
Markov chain relaxes in its quasi-stationary distribution and does not
reach the absorbing state during the finite number of generations. As an example,
for a network with $100$ patches, we present two typical cases in Figure \ref{figevolution}:
when extinction is likely in $100$ generations (replications in solid black lines) and when a quasi-equilibrium is reached 
(replications in broken grey lines).
If the simulations are run long enough, the quasi-equilibrium 
will be left and the system will converge to the coffin state.
% \begin{figure}
%  \begin{center}
%   \includegraphics[width=.5\textwidth]{../sim100c1e008.pdf}
%  \includegraphics[width=.5\textwidth]{../simuQS.pdf}
%  \end{center}
% \caption{Number of occupied patches for two simulations over $100$ generations}
% \label{figevolution}
%  \end{figure}

%%%%%%%%%%%%%%%%%%%%%%%%%%%%%%%%%%%%%%%%%%%%%%%%%%%%%%%%%%%%%%%%%%%%%%%%
\subsubsection{Finite horizon study}
\begin{figure}[h!]
		    \subfloat[][]{\label{figprobaextinc_a}\includegraphics[scale=0.15]{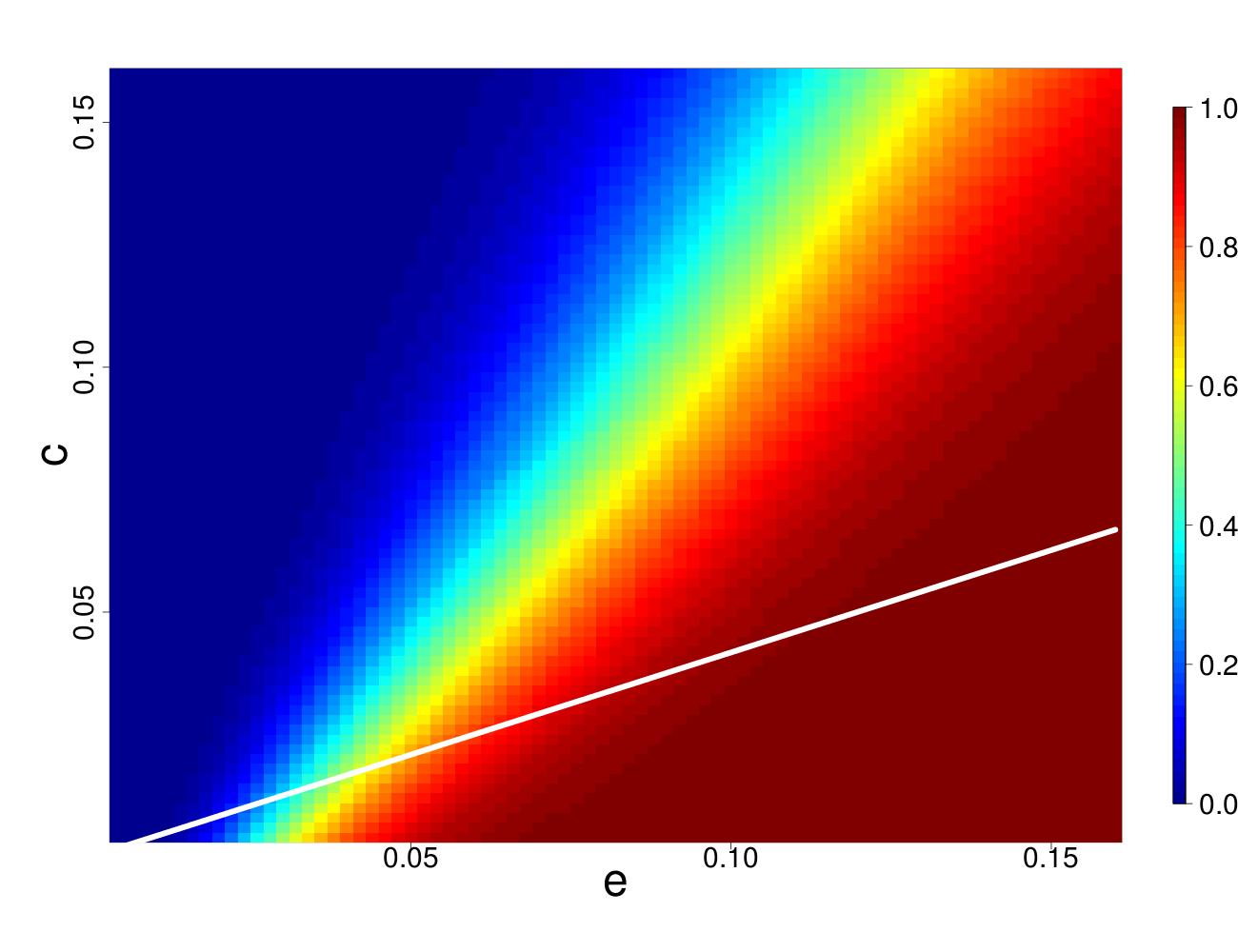}}
		    \subfloat[][]{\label{figprobaextinc_b}\includegraphics[scale=0.15]{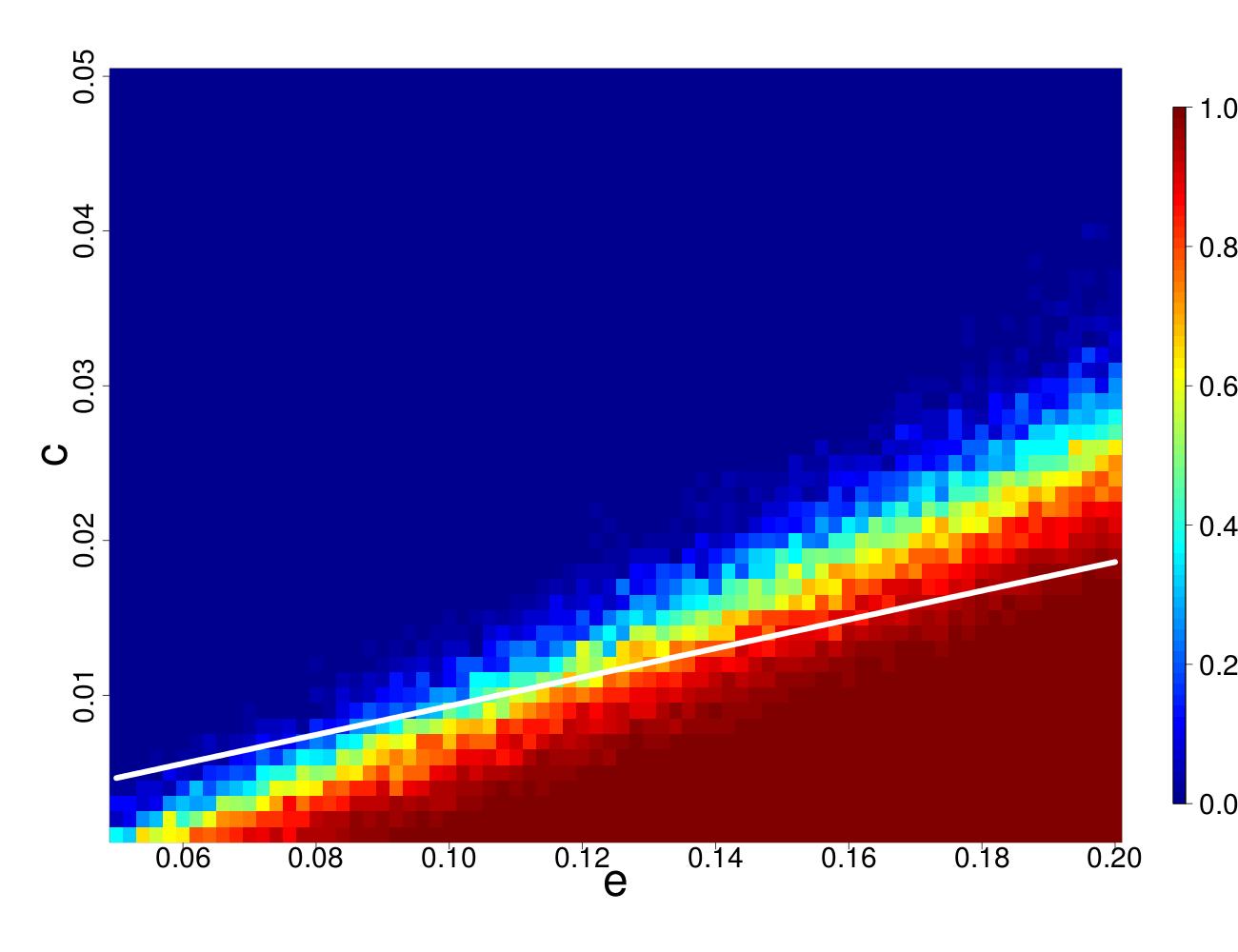}}
	\caption{Probabilities of extinction in $100$ generations for varying values of $e$ and $c$: \protect\subref{figprobaextinc_a} 10 nodes,\protect\subref{figprobaextinc_b} 100 nodes.}
\end{figure}

In a stochastic model, an extinction threshold does not make sense. We
advocate focusing on quantities such as the extinction probability
in a given realistic number of generations
and the mean number of occupied
patches at this generation. 
We aim to study the impact of the network
on persistence through its impact on these two quantities.
Moreover, the impact of $e$ and $c$ must also be taken into account and not only through the ratio $e/c$. 
Indeed, two settings with the same
ratio $e/c$ lead to very different results in a stochastic model according to the order of
magnitude of $e$ and $c$.  
For a fixed network with $10$ nodes and for a fixed network with $100$ nodes, we computed (exactly with $10$ nodes, estimated with $100$ nodes)
the probability of extinction in $100$ generations $\P_{z_0}(T_0\le 100)=\P_{z_0}(\#Z_{100}=0)$
for different values of $e$ and $c$. Here, the initial state $z_0$ is chosen such that all patches are occupied.
The color maps of these probabilities are displayed in Figure \ref{figprobaextinc_a} and \ref{figprobaextinc_b}.

The white line corresponds to the level $e/c=\lambda_{A,1}$ which is the frontier obtained when using the large network approximation.
As observed in this case of a finite network, this line fails to separate cases with a high probability of extinction
from the others. Furthermore, as we want to take the stochasticity of the model due to a finite number of patches into account,
a threshold would not be relevant. Actually, there exists a fuzzy band where there is very little confidence in the behavior of the system. \\

From these remarks, we decided to conduct finite horizon analyses in the following sections. 
The time horizon was chosen with respect to the application.
Both, the probability of persistence and the mean number of occupied patches were studied
to quantify the impact of the network topology in different settings depending on the values of the parameters 
$e$ and $c$.
The next section presents methods for simulation when the probability of persistence is hard to compute.

%% file: analysesensi.tex
The aim was to conduct a sensitivity analysis based on the chosen outputs of the stochastic model (mean number of occupied patches and the 
persistence probability at generation $100$: $\E(\#Z_{100})$ and $\P(\#Z_{100}>0)$) with respect to its parameters: the extinction rate $e$, the colonisation rate $c$ and the 
graph $G$. The graph is parameterised by its density $d$ which is equivalent to the number of edges or the mean degree and its topology i.e. 
the distribution of the edges given a total number of edges.
The variation range of the parameters was fixed in order to induce a context of weak, middle and strong extinction 
since the aim is to study the impact of the network topology
in contrasted situations (see Table \ref{tab10farms} for the values used for the full factorial design of experiments).
The analyses were done for two cases for $n=10$ patches and $n=100$ patches.
For $n=10$, exact computations were still achievable while for $n=100$, the simulation methods  
presented in \ref{subsecMethodsimu} were used. We ensured that the simulations had reached a sufficient degree of precision 
to consider that the part of variability in the outputs due to the estimation method was negligible in comparison with the variation due 
to the input parameter variation.    
Five kinds of networks were compared: an Erd\H{o}s-Rényi network, a community network, a ``lattice'' network, and two preferential attachment
networks with powers $1$ and $3$.
For the community model, only one community setting was studied and the ratio of intra versus inter connection probability 
was set at $100$. When $n=10$, the patches were equally split into two communities. When $n=100$, the patches were equally split into five communities.
For each network structure, ten replicate networks were built with randomly generated edges.

 \begin{table}[h!]
 \begin{center}
  \begin{tabular}{|c|c|c|}
\hline
&10 patches&100 patches\\
\hline
  e & $\{0.05,\ 0.10,\ 0.15 \}$&$\{0.10,\ 0.20,\ 0.25\}$\\
  \hline
   c & $\{0.01,\ 0.05,\ 0.10\}$&$\{0.001,\ 0.005,\ 0.010\}$ \\
   \hline
   $d$ &$\{30\%,\ 50\%,\ 70\% \}$&$\{5\%,\ 10\%,\ 30\% \}$ \\
   \hline
  \end{tabular}
\caption{Values for exploration of the model with $10$ patches or $100$ patches}
\label{tab10farms}
 \end{center}
\end{table}

The sensitivity analyses were based on an analysis of variance. 
The influence of the parameters and their interaction on
$\P(\#Z_{100}>0)$ (actually the logit of this probability)
and on $\E(\#Z_{100})$ were assessed.
The only source of variability was the randomness in the graph generation. We recall that for $n=10$
the computations are exact and for $n=100$, the estimates are precise enough to ensure the significance. %suffisant ?
%neglect their variations.
All these linear models had a coefficient of determination $R^2$ greater than $99.9\%$.
As expected, the parameters $e$, $c$ and $d$ were by far the most important
since a large range of variation was explored for each of these parameters
and since any of these parameters could drive the system to extreme situations where extinction is likely or rare. 
Nevertheless, the topology was still significant. 
As suggested by the significance of high order interactions, especially the ones involving topology, 
a topology was not found to be uniformly (whatever the values of $e$, $c$ or $d$) better (according to
$\P(\#Z_{100}>0)$ or to $\E(\#Z_{100})$) than the others.
\\

The comparison based on $\E(\#Z_{100})$ has shown an inversion in the ranking of the topologies which
was similar to the one noticed by \citet{Gilarranz201211}.
This inversion appeared with both $n=10$ and $n=100$ patches.
%Here, the ranking was based on the mean number of occupied patches at generation $100$: $\E(\#Z_{100})$.
When the combination of values of $e$, $c$ and $d$ ensured persistence with a high probability,
the best topologies % according to $\E(\#Z_{100})$
were those with a better balance in degree distribution such as the lattice, ER and community topologies. 
However, although the difference in mean was found significant, the order of magnitude of this difference was only of a few patches ($\approx 5 $)
for $n=100$ patches. 
On the other hand, the topologies leading to some very connected patches (hub) such as the preferential attachment topologies (especially
when the power parameter is set at $3$) maximized the number of occupied patches when the persistence in the system is threatened in $100$ generations.
In that case, the differences were more contrasted between topologies.
\\

When the topologies were compared according to $\P(\#Z_{100}>0)$,
the same kind of inversion was noticed for $n=10$ patches.
However, the balanced topologies (lattice, ER and community) were found to be better in settings where 
the probability of persistence is greater than $99.5\%$. In other cases when 
the persistence was more jeopardized, the preferential attachment topologies were the best. 
With $n=100$ patches, we have only observed
the better resistance of preferential attachment topologies
and also their crucial role in critical situations where persistence and extinction
had pretty much the same probability of occurring.
However, we were not able to obtain settings where the persistence probability
was greater for any of the balanced topologies than for the preferential attachment topology, even though we have computed it
with Algorithms 2 and 3 for settings where the order of magnitude of the persistence probability is $1-10^{-15}$.
\\

\begin{figure}[h!]
		    \subfloat[][]{\label{figproba}\includegraphics[scale=0.4]{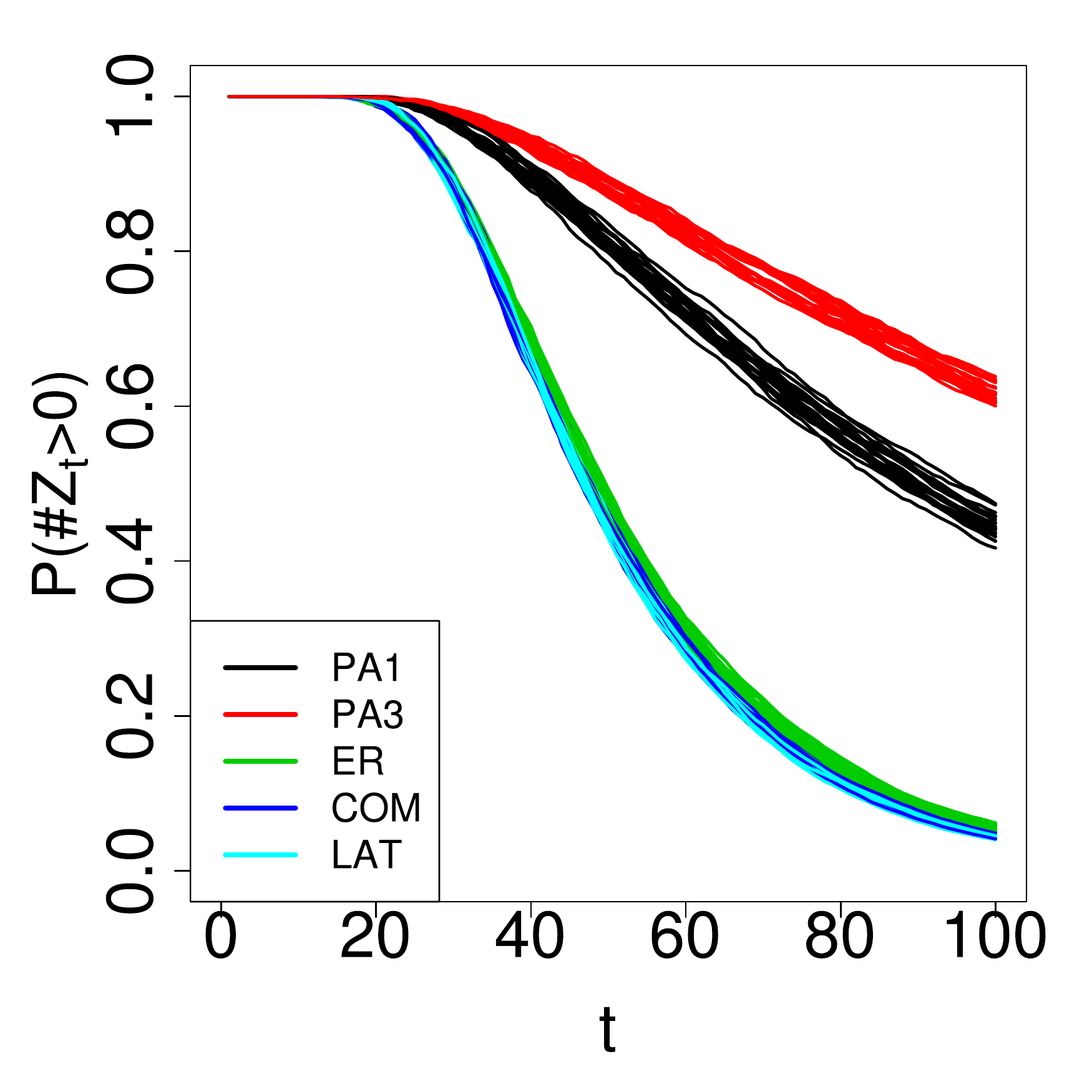}}
		    \subfloat[][]{\label{figesp}\includegraphics[scale=0.4]{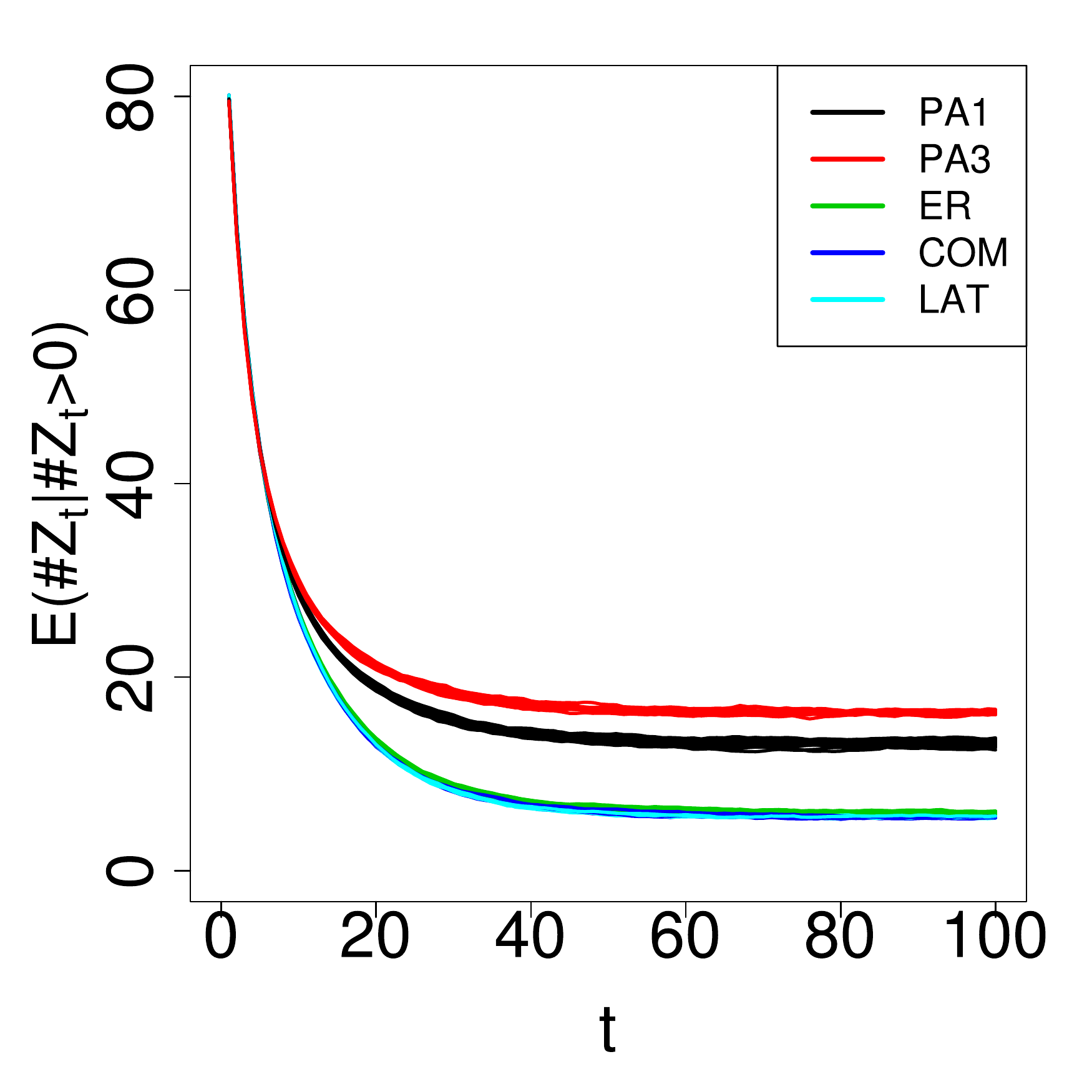}}
	\caption{\protect\subref{figproba} Probability of persistence and \protect\subref{figesp} mean number of occupied patches, 
in varying $t$ generations (based on $20$ replications of the network for a given topology) for $n=100$, $c=0.01$, $e=0.25$ and $d=30\%$.
COM: community network, ER: Erd\H{o}s-Rényi network, LAT: Lattice network, PA1: preferential attachment network with power 1, PA3: preferential attachment with power 3.}
\end{figure}

Figure \ref{figproba} illustrates the crucial role of the topology in the chosen setting. The preferential attachment topology with power $3$ ensured the 
persistence probability in $100$ generations to be greater than $0.6$ while the ER, lattice and community topologies have led to a probability 
of persistence smaller 
than $0.05$. 
Figure \ref{figesp} displays the mean number of occupied patches conditionally to persistence. This figure
shows that the quasi-stationary distributions for the preferential attachment topologies have led to mean numbers of
occupied patches which were close to $15$ while these conditional mean numbers were close to $5$ for the other three topologies.
Since the number of occupied patches was too close to $0$ in this quasi-equilibrium,
the system had a very low probability of 
relaxing for a while in its quasi-stationary distribution.
\\

\begin{figure}[ht]
		    \subfloat[][]{\label{100exemple_a}\includegraphics[scale=0.4]{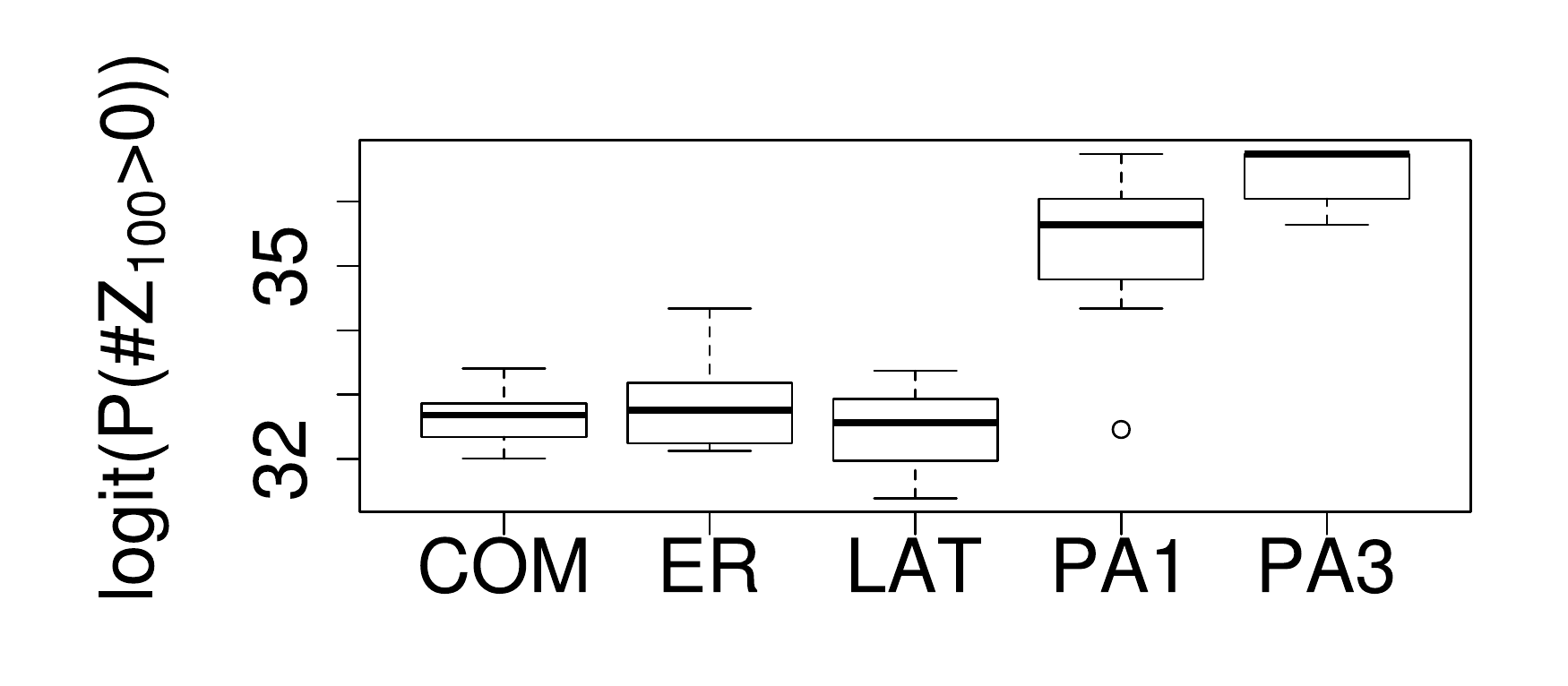}}
		    \subfloat[][]{\label{100exemple_b}\includegraphics[scale=0.4]{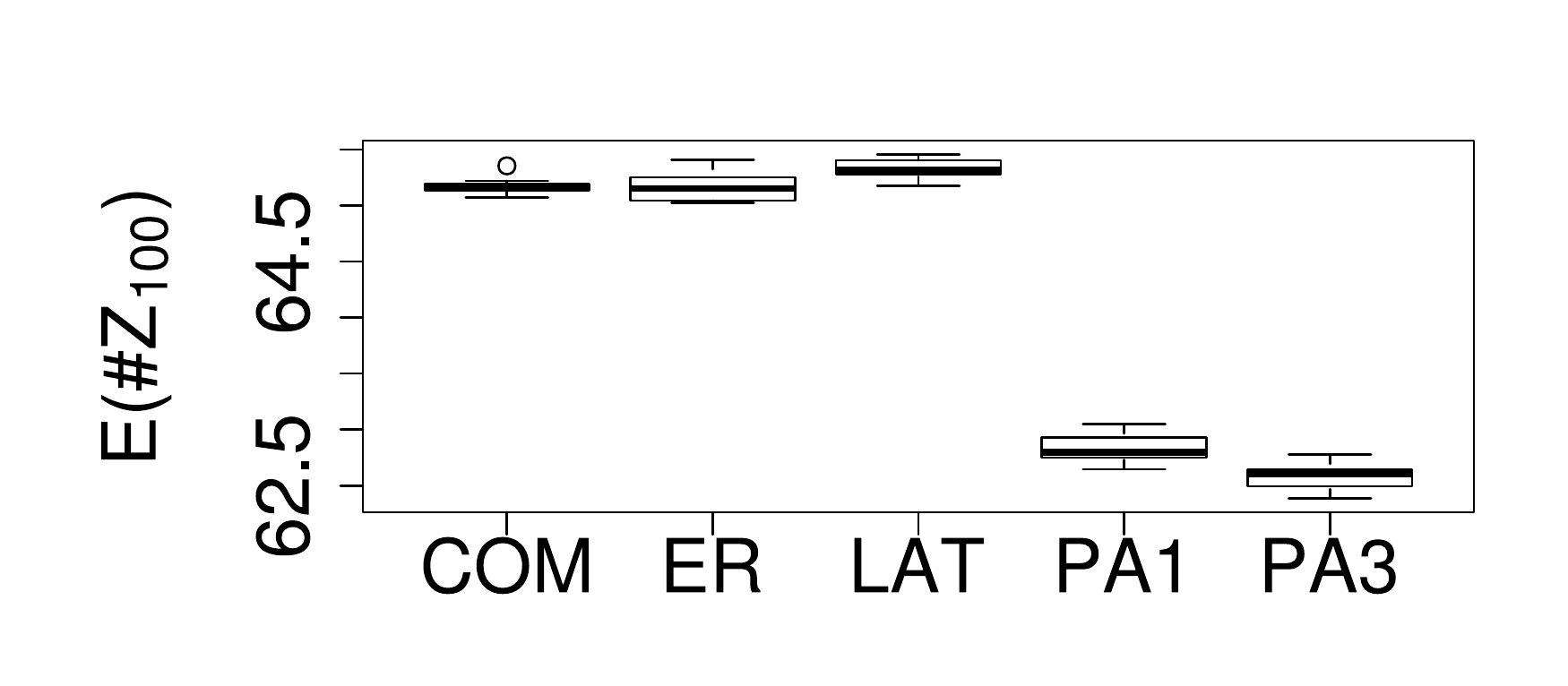}}
	\caption{Boxplots of the probabilities of persistence over $100$ generations \protect\subref{100exemple_a} and the number of occupied patches at generation $100$ 
	\protect\subref{100exemple_b}, computed with 10 replications of each network topology. COM: community network, ER: Erd\H{o}s-Rényi network, LAT: Lattice network, PA1: preferential attachment network with power 1, PA3: preferential attachment  with power 3.}
\label{100exempleaetb}
	\end{figure}

\paragraph{Conclusion}
From this study, we have determined that the role of the topology is not always crucial. But in some settings,
it has a key impact on persistence probability and thus on the mean number of occupied patches.
Thanks to sharp computations of the persistence probabilities, the differences between the topologies were also 
highlighted on the basis of rare events (rare persistence or rare extinction). 
The preferential attachment topology is generally more resistant according to this probability
especially when the persistence is jeopardized.
% In this sensitivity study, the preferential attachment topology ensured a bigger probability of persistence than did the other topologies
% if the order of magnitude of this probability was bigger than
% $0.05\%$ for $n=10$ patches.
% For $n=100$ the preferential attachment topology was always better according to the probability of persistence
% even when it was around $1-10^{-15}$.
Nevertheless, concerning the occupancies, balanced topologies can perform a little better even though 
they still have a smaller probability of persistence than do the preferential attachment topologies.
As an example, Figure \ref{100exemple_a} and \ref{100exemple_b}
%\ref{100exempleaetb}
display $\P(\#Z_{100}>0)$ and $\E(\#Z_{100})$ in a particular setting.
Each of these two criteria may rank the settings including the topology in different orders.

The community topology may be a little more sensitive
to extinction than are ER or lattice topologies but they are globally equivalent for this dynamic model.
%In our study, there were $2$ groups for $n=10$ and $5$ for $n=100$ which ensured that the groups were not too small.
%demander ce qui se passerait si c'était plus petit
Even if it is quite obvious, we mention that it was noticed that the role of the topology is enhanced when the density is higher, when
$c$ is greater and when the number of patches is greater. 

%% file: art_20140324arxiv.bbl
\begin{thebibliography}{}

\bibitem[Adler and Nuernberger, 1994]{Adler1994}
Adler, F. and Nuernberger, B. (1994).
\newblock Persistence in patchy irregular landscapes.
\newblock {\em Theoretical Population Biology}, 45(1):41 -- 75.

\bibitem[Albert and Barab\'asi, 2002]{Albert2002}
Albert, R. and Barab\'asi, A.-L. (2002).
\newblock Statistical mechanics of complex networks.
\newblock {\em Reviews of Modern Physics}, 74(1):47--97.

\bibitem[Amrein and K\"{u}nsch, 2011]{amrein2011}
Amrein, M. and K\"{u}nsch, H.~R. (2011).
\newblock A variant of importance splitting for rare event estimation: Fixed
  number of successes.
\newblock {\em ACM Transactions on Modeling and Computer Simulation},
  21(2):13:1--13:20.

\bibitem[Barab\'{a}si and Albert, 1999]{barabasi1999}
Barab\'{a}si, A.-L. and Albert, R. (1999).
\newblock Emergence of scaling in random networks.
\newblock {\em Science}, 286(5439):509--512.

\bibitem[Bocci and Chable, 2008]{bocci_peasant_2008}
Bocci, R. and Chable, V. (2008).
\newblock Peasant seeds in europe: Stakes and prospects.
\newblock {\em Cahiers Agricultures}, 17(2):216--21.

\bibitem[Chakrabarti et~al., 2008]{Chakrabarti2008}
Chakrabarti, D., Wang, Y., Wang, C., Leskovec, J., and Faloutsos, C. (2008).
\newblock Epidemic thresholds in real networks.
\newblock {\em ACM Transactions on Information and System Security},
  10(4):1:1--1:26.

\bibitem[Csardi and Nepusz, 2006]{igraph}
Csardi, G. and Nepusz, T. (2006).
\newblock The igraph software package for complex network research.
\newblock {\em InterJournal}, Complex Systems:1695.

\bibitem[Darroch and Seneta, 1965]{Darroch1965}
Darroch, J.~N. and Seneta, E. (1965).
\newblock {On quasi-stationary distributions in absorbing discrete-time finite
  Markov chains}.
\newblock {\em Journal of Applied Probability}, 2(1):88--100.

\bibitem[Day and Possingham, 1995]{Day1995}
Day, J.~R. and Possingham, H.~P. (1995).
\newblock A stochastic metapopulation model with variability in patch size and
  position.
\newblock {\em Theoretical Population Biology}, 48:333--360.

\bibitem[Del~Moral and Doucet, 2009]{delmoraldoucet2009}
Del~Moral, P. and Doucet, A. (2009).
\newblock Particle methods: An introduction with applications.
\newblock Rapport de recherche RR-6991, INRIA.

\bibitem[Demeulenaere and Bonneuil, 2011]{demeulenaere_semences_2012}
Demeulenaere, E. and Bonneuil, C. (2011).
\newblock {Des Semences en partage : construction sociale et identitaire d'un
  collectif "paysan" autour de pratiques semenci{\`e}res alternatives.}
\newblock {\em Techniques \& Culture}, 57(2011/2):202--221.

\bibitem[Demeulenaere et~al., 2008]{demeulenaere_etude_2008}
Demeulenaere, E., Bonneuil, C., Balfourier, F., Basson, A., Berthellot, J.,
  Chesneau, V., Ferté, H., Galic, N., Kastler, G., Koening, J., Mercier, F.,
  Payement, J., Pommart, A., Ronot, B., Rousselle, Y., Supiot, N., Zaharia, H.,
  and Goldringer, I. (2008).
\newblock \'etude des complémentarités entre gestion dynamique à la ferme et
  gestion statique en collection: cas de la variété de blé {Rouge de
  Bordeaux}.
\newblock {\em Les Actes du {BRG}}, 7:117-- 138.

\bibitem[Erd\H{o}s and R\'{e}nyi, 1959]{erdosrenyi1959}
Erd\H{o}s, P. and R\'{e}nyi, A. (1959).
\newblock On random graphs, {I}.
\newblock {\em Publicationes Mathematicae Debrecen}, 6:290--297.

\bibitem[Franc, 2004]{Franc2004}
Franc, A. (2004).
\newblock Metapopulation dynamics as a contact process on a graph.
\newblock {\em Ecological Complexity}, 1(1):49 -- 63.

\bibitem[Gilarranz and Bascompte, 2012]{Gilarranz201211}
Gilarranz, L.~J. and Bascompte, J. (2012).
\newblock Spatial network structure and metapopulation persistence.
\newblock {\em Journal of Theoretical Biology}, 297(0):11 -- 16.

\bibitem[Hanski and Ovaskainen, 2000]{Hanski2000}
Hanski, I. and Ovaskainen, O. (2000).
\newblock {The metapopulation capacity of a fragmented landscape}.
\newblock {\em Nature}, 404:755--758.

\bibitem[Levins, 1969]{Levins1969}
Levins, R. (1969).
\newblock Some demographic and genetic consequences of environmental
  heterogeneity for biological control.
\newblock {\em Bulletin of the ESA}, pages 237--240.

\bibitem[M\'el\'eard and Villemonais, 2012]{Meleard2012}
M\'el\'eard, S. and Villemonais, D. (2012).
\newblock Quasi-stationary distributions and population processes.
\newblock {\em Probability Surveys}, 9:340--410.

\bibitem[Nowicki and Snijders, 2001]{nowickiSnijders2001}
Nowicki, K. and Snijders, T. A.~B. (2001).
\newblock Estimation and prediction for stochastic blockstructures.
\newblock {\em Journal of the American Statistical Association},
  96(455):1077--1087.

\bibitem[Pautasso et~al., 2013]{pautasso_seed_2013}
Pautasso, M., Aistara, G., Barnaud, A., Caillon, S., Clouvel, P., Coomes,
  O.~T., Delêtre, M., Demeulenaere, E., Santis, P.~D., Döring, T., Eloy, L.,
  Emperaire, L., Garine, E., Goldringer, I., Jarvis, D., Joly, H.~I., Leclerc,
  C., Louafi, S., Martin, P., Massol, F., {McGuire}, S., {McKey}, D., Padoch,
  C., Soler, C., Thomas, M., and Tramontini, S. (2013).
\newblock Seed exchange networks for agrobiodiversity conservation. {A} review.
\newblock {\em Agronomy for Sustainable Development}, 33(1):151--175.

\bibitem[Peyrard et~al., 2008]{Peyrard2008}
Peyrard, N., Dieckmann, U., and Franc, A. (2008).
\newblock Long-range correlations improve understanding of the influence of
  network structure on contact dynamics.
\newblock {\em Theoretical Population Biology}, 73(3):383--394.

\bibitem[Read et~al., 2008]{read_dynamic_2008}
Read, J.~M., Eames, K. T.~D., and Edmunds, W.~J. (2008).
\newblock Dynamic social networks and the implications for the spread of
  infectious disease.
\newblock {\em Journal of The Royal Society Interface}, 5(26):1001--1007.

\bibitem[Rubino and Tuffin, 2009]{rubinotuffin2009}
Rubino, G. and Tuffin, B., editors (2009).
\newblock {\em Rare event simulation using {Monte Carlo} methods}.
\newblock Wiley.

\bibitem[Solé and Bascompte, 2006]{sole_self-organization_2006}
Solé, R.~V. and Bascompte, J. (2006).
\newblock {\em Self-Organization in Complex Ecosystems.}
\newblock Princeton University Press.

\bibitem[Thomas et~al., 2011]{thomas_seed_2011}
Thomas, M., Dawson, J.~C., Goldringer, I., and Bonneuil, C. (2011).
\newblock Seed exchanges, a key to analyze crop diversity dynamics in
  farmer-led on-farm conservation.
\newblock {\em Genetic Resources and Crop Evolution}, 58(3):321--338.

\bibitem[Thomas et~al., 2012]{thomas_-farm_2012}
Thomas, M., Demeulenaere, E., Dawson, J.~C., Khan, A.~R., Galic, N.,
  Jouanne-Pin, S., Remoue, C., Bonneuil, C., and Goldringer, I. (2012).
\newblock On-farm dynamic management of genetic diversity: the impact of seed
  diffusions and seed saving practices on a population-variety of bread wheat.
\newblock {\em Evolutionary Applications}, 5(8):779–795.

\bibitem[van Doorn and Pollett, 2009]{Vandoorn2009}
van Doorn, E. and Pollett, P. (2009).
\newblock Quasi-stationary distributions for reducible absorbing {Markov}
  chains in discrete time.
\newblock {\em Markov Processes and Related Fields}, 15(2):191--204.

\end{thebibliography}
